\documentclass[12pt]{article}
\DeclareMathAlphabet{\scr}{U}{rsfs}{m}{n}
\usepackage{latexsym}
\usepackage[mathscr]{eucal}
\usepackage{amsfonts}
\usepackage{amscd}
\usepackage{cite}
\usepackage{array}
\usepackage{amssymb}
\usepackage{colordvi}
\usepackage[centertags]{amsmath}
\usepackage{enumerate}
\usepackage{graphicx}
\usepackage{booktabs}
\usepackage{theorem}
\usepackage[footnotesize]{caption}
\usepackage{multirow}
\usepackage{color}
\usepackage{dsfont}
\usepackage{dsfont}
\usepackage{amsmath}
\usepackage{pifont}
\usepackage{bm}
\usepackage{ulem}

\newcommand{\cleqn}{\setcounter{equation}{0}}
\newcommand{\newc}{\newcommand}
\newc{\be}{\begin{equation}}
\newc{\ee}{\end{equation}}
\newc{\bea}{\begin{eqnarray}}
\newc{\eea}{\end{eqnarray}}
\newc{\ol}{\overline}
\newc{\wt}{\widetilde}
\newc{\bs}{\boldsymbol}
\newc{\m}{\mathcal}

\begin{document}

\title{\hfill ~\\[0mm]
       \textbf{Spontaneous CP violation from vacuum alignment in $\bs{S_4}$ models of leptons}        }
\date{}
\author{\\[1mm]Gui-Jun Ding$^{1\,}$\footnote{E-mail: {\tt dinggj@ustc.edu.cn}}~,~
Stephen F. King$^{2\,}$\footnote{E-mail: {\tt king@soton.ac.uk}}~,~
Christoph Luhn$^{3\,}$\footnote{E-mail: {\tt christoph.luhn@durham.ac.uk}}~,\\[2mm]
Alexander J. Stuart$^{2\,}$\footnote{E-mail: {\tt a.stuart@soton.ac.uk}}\\ \\[1mm]
  \it{\small $^1$Department of Modern Physics, University of Science and
    Technology of China,}\\
  \it{\small Hefei, Anhui 230026, China}\\[4mm]
  \it{\small $^2$School of Physics and Astronomy, University of Southampton,}\\
  \it{\small Southampton, SO17 1BJ, United Kingdom}\\[4mm]
  \it{\small $^3$IPPP, Department of Physics, University of Durham,}\\
  \it{\small Durham, DH1 3LE, United Kingdom}}

\maketitle

\begin{abstract}
\noindent
We construct models of leptons based on $S_4$ family symmetry combined with a generalised CP symmetry $H_{CP}$. We show how the flavon potential can spontaneously break the symmetry $S_4 \rtimes H_{CP}$ down to $Z_2 \times H^{\nu}_{CP}$ in the neutrino sector, where the choice of preserved CP symmetry $H^{\nu}_{CP}$ is controlled by free (real) parameters in the flavon potential. We propose two realistic models of this kind, one at the effective level and one at the renormalisable level. Both models predict trimaximal lepton mixing with CP being either fully preserved or maximally broken, with the intermediate possibility forbidden by the structure of the models.
\end{abstract}

\thispagestyle{empty}
\vfill
\newpage
\setcounter{page}{1}

\section{\label{sec:1}Introduction}
\cleqn

Recent hints~\cite{Abe:2011sj,Adamson:2011qu,Abe:2011fz} for a nonzero reactor mixing angle $\theta_{13}$, and its subsequent measurement by the Daya Bay~\cite{DayaBay}, RENO~\cite{RENO} and Double Chooz~\cite{DCt13}
Collaborations have revealed an unexpectedly large mixing angle of about
$9^\circ$ to an accuracy comparable to the other two lepton mixing angles,
i.e. the solar and atmospheric angles~\cite{pdg}. This observation has had a major impact on models of leptons based on discrete family symmetry (for a recent review see e.g. \cite{King:2013eh}). On the experimental side, this remarkable discovery opens up the possibility of unravelling (some of) the remaining unknown parameters of the lepton sector, the neutrino mass ordering as well as the leptonic CP violating phases,\footnote{The leptonic CP phases include one Dirac phase $\delta_{CP}$
and two Majorana phases $\alpha_{21}$, $\alpha_{31}$. The former can be
measured in neutrino oscillation experiments, the latter are relevant for
neutrinoless double beta decay and more difficult to be measured.} which is
the primary goal of next generation neutrino oscillation experiments.

Concerning the Dirac CP violating phase $\delta_{CP}$, the latest global
analyses of the available neutrino oscillation data do not allow to pin down a preferred value at the 3$\sigma$ level~\cite{Tortola:2012te,Fogli:2012ua,GonzalezGarcia:2012sz}. Yet, as we
move into the age of precision measurement of the lepton mixing parameters,
this state of affairs is bound to change. Inspired by the successes of discrete family symmetries in predicting the lepton mixing angles, it is not unreasonable to believe that the symmetry approach can also lead to predictions for $\delta_{CP}$. To find a realisation of this idea, it appears natural to combine a family symmetry and generalised CP symmetries such that both the mixing angles and the CP violating phases are predicted simultaneously. In this setup, the breaking of the family symmetry would give rise to particular mixing angles, while the breaking of the generalised CP symmetries leads to particular values of the CP phases.

The concept of generalised CP transformations has been around for
decades, and it was realised early on that there exists a one-to-one
correspondence between the generalised CP transformations and the automorphism group of an imposed family symmetry~\cite{Ecker:1981wv,Ecker:1983hz,Ecker:1987qp,Neufeld:1987wa,Grimus:1995zi}.
Recently, the consistent generalised CP transformations (i.e. the automorphism group) of the discrete family symmetry groups of order smaller than 31, with irreducible triplet representations, have been discussed in~\cite{Holthausen:2012dk}.

Despite these fundamental studies, there has been only little work on
constructing explicit models which realise the idea of combining family and
generalised CP symmetries \cite{Harrison:2002et,Mohapatra:2012tb,Feruglio:2012cw}. In Ref.~\cite{Mohapatra:2012tb}, an $S_4$ model with imposed generalised CP symmetry is presented which, adopting the type II seesaw mechanism, predicts a normal neutrino mass ordering, a maximal atmospheric angle and a maximal Dirac CP violating phase, $\delta_{CP}=\pm\pi/2$. Based on pure symmetry arguments, a systematic classification of an $S_4$ family symmetry combined with generalised CP symmetries is performed in ~\cite{Feruglio:2012cw}. Although the resulting phenomenology was analysed in great detail, a dynamical model which implemented these ideas was not given in~\cite{Feruglio:2012cw}.

In this paper we shall address the question of breaking a generalised CP symmetry spontaneously in the framework of models based on $S_4$ family symmetry. We shall construct models of leptons imposing both $S_4$ family symmetry and the corresponding generalised CP symmetry $H_{CP}$.
We shall then show how the flavon potential can spontaneously break the symmetry $S_4 \rtimes H_{CP}$ down to $Z_2 \times H^{\nu}_{CP}$ in the neutrino sector, which was simply assumed to happen in~\cite{Feruglio:2012cw}. In our models, the choice of preserved CP symmetry $H^{\nu}_{CP}$ in the neutrino sector is controlled by free (real) parameters in the flavon potential. We propose two realistic models of this kind, one at the effective level and one at the renormalisable level. Both models predict trimaximal lepton mixing~\cite{TMref,King:2011zj} with CP being either fully preserved or maximally broken, with the intermediate possibility forbidden by the structure of the models.

The layout of the rest of this paper is as follows. In section~2 we draw
together some of the basic observations about generalised CP symmetry when
combined with a discrete family symmetry that are rapidly becoming to be
appreciated in the literature. In section~3 we focus on the case of $S_4$ family symmetry, which was also analysed extensively in~\cite{Feruglio:2012cw}. However, in our case, we will work in a different basis, the diagonal charged lepton basis, where the representation matrix for the order three generators $T$ is diagonal, and we verify explicitly that the same physical results emerge, as expected. This discussion also sets the notation and allows us to be rather brief when discussing the physical implications of the models which follow. In section~4
we present our first example of an $S_4 \rtimes H_{CP}$ model at the effective level (i.e. involving non-renormalisable operators) where the flavon potential leads to exactly the kinds of possibilities for CP violation discussed in section~3. In section~5 we present a superior renormalisable $S_4 \rtimes H_{CP}$ model which has the added advantage of explaining why the reactor angle is smaller than the other lepton mixing angles, and also leads to similar options for CP violation. Section~6 concludes the main body of the paper, followed by two appendices on more technical aspects.

\section{\label{sec:2}A consistent definition of generalised CP transformations}

In this section, we start by briefly reviewing for completeness how CP
transformations are generally defined in a consistent way {\it without} an existing family symmetry. This is begun by noting that in the extension of the Standard Model (SM) with Majorana masses for the left-handed neutrinos, the mass terms and the charged current interactions for the lepton fields read (below the electroweak symmetry breaking scale),
\begin{equation}
\label{eq:lag}\mathcal{L}=-\bar{l}_Lm_ll_R-\frac{1}{2}\nu^{T}_LCm_{\nu}\nu_L+\frac{g}{\sqrt{2}}\,\bar{l}_L\gamma^{\mu}\nu_LW^{-}_{\mu}+h.c.
\end{equation}
$l_{L,R} \equiv (e, \mu, \tau)_{L,R}^T$ stands for the SM charged lepton
fields, $\nu_{L} \equiv (\nu_e, \nu_\mu, \nu_\tau)_{L}^T$ are the left-handed neutrino fields, $C$ is the charge conjugation matrix, $m_{l,\nu}$ are complex matrices, and $m_{\nu}$ is symmetric. The charged current interactions are invariant under the so-called generalised CP transformation defined by\footnote{A factor of $i$ is included,
differing from the definition of Ref.\cite{Branco:2011zb}, in order to eliminate an extraneous ``$-$'' sign from the second equation of Eq.~(\ref{eq:gc}).}
\begin{eqnarray}
\begin{gathered}
 \, l_L(x)\,  \overset{CP}{\rightarrow} i X_L \gamma^0 C\, \bar{l}_{L}^{\,T}(x'),\\
 \, \nu_L(x)\,\overset{CP}{\rightarrow}i X_L \gamma^0 C\, \bar{\nu}_{L}^{\,T}(x'),\\
 \, l_R(x)\,  \overset{CP}{\rightarrow} i X_R \gamma^0 C\, \bar{l}_{R}^{\,T}(x'),
\end{gathered}
\end{eqnarray}
where $X_L$ and $X_R$ are unitary matrices acting on generation space and
$x'=(t,-\bf{x})$. Notice that the ``canonical''/traditional CP transformation can be recovered from the above equations by setting both $X_L$ and $X_R$ to the identity matrix. Furthermore, note that the lepton fields $l_L$ and $\nu_L$ have to transform in the same way due to the presence of the left-handed charged-current interactions. Then, the Lagrangian of Eq.~(\ref{eq:lag}) conserves CP if and only if the lepton mass matrices $m_{l}$ and $m_{\nu}$ satisfy the following relations,
\begin{equation}
\label{eq:gc}X_L^{\dagger}m_{l}X_R =m^{*}_{l},\qquad X^{T}_L m_{\nu}X_L=m^{*}_{\nu}.
\end{equation}

Let us now consider a theory that is invariant under both a generalised CP
symmetry {\it and} a family symmetry $G_f$, which is the main focus of this
paper. We include this discussion in order to set the notation and to make the rest of the paper accessible. However, since all the information in this
section is already in the literature, the discussion below is necessarily
brief and we refer interested readers to e.g.~\cite{Feruglio:2012cw,Holthausen:2012dk} for more details.
Let us, then, consider a field $\varphi$ in a generic irreducible
representation $\mathbf{r}$ of $G_f$ which transforms under the
action of $G_f$ as
\begin{equation}
\varphi(x)\stackrel{G_f}{\longrightarrow} \rho_{\mathbf{r}}(g) \varphi(x),
\qquad g \in G_f\ ,
\end{equation}
where $\rho_{\mathbf{r}}(g)$ denotes the representation matrix for the element $g$ in the irreducible representation $\mathbf{r}$. The mapping of $\varphi$ under a generalised CP transformation is given
by~\cite{Ecker:1981wv,Ecker:1983hz,Ecker:1987qp,Branco:2011zb,Feruglio:2012cw,Holthausen:2012dk}:
\begin{equation}
\varphi(x)\stackrel{CP}{\longrightarrow}X_{\mathbf{r}}\,\varphi^{*}(x')\ ,
\end{equation}
where $X_{\mathbf{r}}$ is a unitary matrix in order to
keep the kinetic term invariant, and the obvious action of CP on the spinor
indices has been omitted for the case of $\varphi$ being a spinor.

Requiring a generalised CP symmetry $H_{CP}$ in the context of a flavour model based on some family symmetry $G_f$, restricts the allowed choices for $X_{\bf{r}}$ considerably~\cite{Feruglio:2012cw,Holthausen:2012dk}.
If we first perform a CP transformation, followed by a family symmetry
transformation, and subsequently an inverse CP transformation we obtain
\begin{equation}
\varphi(x)\stackrel{CP}{\longrightarrow}X_{\mathbf{r}}\,\varphi^{*}(x')\stackrel{G_f}{\longrightarrow}X_{\mathbf{r}}\rho^{*}_{\mathbf{r}}(g)\varphi^{*}(x')\stackrel{CP^{-1}}{\longrightarrow}X_{\mathbf{r}}\rho^{*}_{\mathbf{r}}(g)X^{-1}_{\mathbf{r}}\varphi(x)\ .
\end{equation}
As the theory should be invariant under this sequence of transformations, the resulting net transformation must be equivalent to a family symmetry
transformation $\rho_{\mathbf{r}}(g')$ of some group element $g'$
\begin{equation}
\label{eq:consistency}X_{\mathbf{r}}\rho^{*}_{\mathbf{r}}(g)X^{-1}_{\mathbf{r}}=\rho_{\mathbf{r}}(g'),
\qquad g,g' \in G_f\ .
\end{equation}

This equation defines the so-called {\it consistency equation} which must be
satisfied in order to implement both generalised CP and family symmetries
simultaneously.  It is important to note that Eq.~(\ref{eq:consistency}) must hold for {\it all} representations ${\bf r}$ simultaneously, i.e. the elements $g$ and $g'$ must be the same for all irreducible representations. Furthermore, Eq.~\eqref{eq:consistency} implies that the generalised CP transformation $X_{\mathbf{r}}$ maps the group element $g$ onto $g'$, and this mapping preserves the family symmetry's group
structure.\footnote{$X_{\mathbf{r}}\rho^{*}_{\mathbf{r}}(g_1g_2)X^{-1}_{\mathbf{r}}=X_{\mathbf{r}}\rho^{*}_{\mathbf{r}}(g_1)\rho^{*}_{\mathbf{r}}(g_2)X^{-1}_{\mathbf{r}}=X_{\mathbf{r}}\rho^{*}_{\mathbf{r}}(g_1)X^{-1}_{\mathbf{r}}
X_{\mathbf{r}}\rho^{*}_{\mathbf{r}}(g_2)X^{-1}_{\mathbf{r}}=\rho_{\mathbf{r}}(g'_1)\rho_{\mathbf{r}}(g'_2)=\rho_{\mathbf{r}}(g'_1g'_2)$, where we denote $X_{\mathbf{r}}\rho^{*}_{\mathbf{r}}(g_i)X^{-1}_{\mathbf{r}}=\rho_{\mathbf{r}}(g'_i)$.
Therefore the CP transformation $X_{\mathbf{r}}$ is a homomorphism of the family symmetry group $G_f$.} For faithful representations~${\mathbf{r}}$ where the function $\rho_{\mathbf{r}}$ maps each element of $G_f$ into a distinct matrix, the consistency equation will define a unique mapping of the abstract group $G_f$ to itself.\footnote{The consistency equation defines an automorphism for the group $G_f$.  See Ref. \cite{Holthausen:2012dk} for a more formal treatment.} It is also clear from Eq.~(\ref{eq:consistency}) that $g$ and $g'$ must be of the same order.

It is important to note that, if $X_{\mathbf{r}}$ is a solution to Eq.~\eqref{eq:consistency}, then not only is $e^{i\theta}X_{\mathbf{r}}$ a solution (with arbitrary phase $\theta$), but also
\be
\rho_{\mathbf{r}}(h)X_{\mathbf{r}},\text{~ ~with~ $h\in G_f$,}
\ee
are solutions as well.\footnote{To see this note that
$\rho_{\mathbf{r}}(h)X_{\mathbf{r}}\rho^{*}_{\mathbf{r}}(g)(\rho_{\mathbf{r}}(h)X_{\mathbf{r}})^{-1}=
\rho_{\mathbf{r}}(h)X_{\mathbf{r}}\rho^{*}_{\mathbf{r}}(g)X^{-1}_{\mathbf{r}}\rho^{-1}_{\mathbf{r}}(h)
=\rho_{\mathbf{r}}(h)\rho_{\mathbf{r}}(g')\rho_{\mathbf{r}}(h^{-1})=\rho_{\mathbf{r}}(hg'h^{-1})$. Therefore the generalised CP transformation $\rho_{\mathbf{r}}(h)X_{\mathbf{r}}$ maps the group element $g$ into $hg'h^{-1}$ which belongs to the conjugacy class of $g'$. It is equivalent to a CP transformation $X_{\mathbf{r}}$ plus an inner automorphism of the family symmetry group $G_f$~\cite{Holthausen:2012dk}.}
Therefore the consistency equation can only determine the possible form of the CP transformation $X_{\mathbf{r}}$ up to an overall arbitrary phase and a $G_f$ transformation $\rho_{\bf r}(h)$ for a given irreducible representation $\mathbf{r}$. All the above statements apply to any family symmetry group $G_f$ regardless of it being discrete or continuous. For discrete family symmetries, it is sufficient to impose the consistency equation Eq.~\eqref{eq:consistency} on the group's generators.

\section{\label{sec:general}$\bs{S_4} \rtimes \bs{H_{CP}}$ scenarios with $\bs{Z_2}  \times \bs{H^\nu_{CP}}$ preserved in neutrino sector and $\bs{Z_3}$ in the charged lepton sector}
\cleqn

In the remainder of the paper, we will investigate the generalised CP transformations $H_{CP}$ consistent with $S_4$ as a family symmetry group,
\be
S_4 \rtimes H_{CP}.
\ee
In other words we shall be interested in theories which (at high energies above the symmetry breaking scale) respects $S_4 \rtimes H_{CP}$ where the fields transform as
\begin{eqnarray}
\varphi(x)&\stackrel{S_4}{\longrightarrow}& \rho_{\mathbf{r}}(g) \varphi(x),
\qquad g \in S_4\ ,\\[3mm]
\varphi(x)&\stackrel{H_{CP}}{\longrightarrow}&X_{\mathbf{r}}\,\varphi^{*}(x')\ ,
\end{eqnarray}
where the elements of $H_{CP}$ are denoted by $X_{\bf r} $. Since $S_4$ is our primary focus, in Appendix~\ref{app2}, we prove explicitly the statement in \cite{Holthausen:2012dk} that the most general CP transformation $H_{CP}$ consistent with the $S_4$ flavour group is simply given by $X_{\bf r}$ being equal to the identity (up to an inner automorphism). Therefore the most general CP transformation $H_{CP}$ consistent with $S_4$ family symmetry is of the same form as the family group transformation itself,
\begin{equation}
\label{24}
X_{\bf r} ~=~\rho_{\bf r} (h) \ , \qquad h\in S_4 \ ,
\end{equation}
where $h$ can be any of the 24 group elements of $S_4$.

Following~\cite{Feruglio:2012cw}, we further assume in this section that
the underlying combined symmetry group $S_4 \rtimes H_{CP}$ is broken to
$G^{\nu}_{CP}$ in the neutrino sector and $G^\ell$ in the charged lepton
sector\footnote{In the models constructed in sections~4 and~5, the $Z_3$
subgroup of $S_4$ will be broken. However, this happens in such a way that
the resulting charged lepton mass matrix remains diagonal.}
\be
G^\nu_{CP} \cong Z_2\times H^\nu_{CP} \ ~\text{and} \qquad G^\ell \cong Z_3 \ .
\ee
The main purpose of this paper (discussed in the following sections) is to show how the original symmetry $S_4 \rtimes H_{CP}$ can be spontaneously broken to $Z_2\times H^\nu_{CP}$ in the neutrino sector, which goes beyond the analysis in~\cite{Feruglio:2012cw}. However, in this section, we begin by following in the footsteps of~\cite{Feruglio:2012cw} but using a different $S_4$ basis, namely that in which the charged leptons are diagonal and all the mixing arises from the neutrino sector. In fact we regard it as necessary to repeat the symmetry analysis in the new basis in order to verify explicitly that the two bases give equivalent results, thereby placing our later results on a sound footing. It is also useful to set the notation and to allow us to arrive at the physical results very efficiently in the next sections. Therefore, although the results in this section may seem like repetition, we regard it as both necessary and useful to proceed by first following the analysis in~\cite{Feruglio:2012cw} in a different basis.

The basis in which we work in this paper is motivated by the trimaximal $S_4$ model of~\cite{King:2011zj} (originally proposed without $H_{CP}$), where the resulting symmetry breaking led to a preserved $Z^S_2$ in the neutrino sector generated by $S$, as well as a diagonal charged lepton mass matrix which arises thanks to the chosen $S_4$ basis with diagonal complex $T$ generator, as discussed in Appendix~\ref{app2}. The $Z_2^S$ symmetry of the neutrino sector will automatically produce a neutrino mass matrix in which the second column of the corresponding mixing matrix is proportional to $(1,1,1)^T$. In the next section we shall discuss models which are inspired by this  model but include $H_{CP}$ and are hence based on $S_4 \rtimes H_{CP}$ which is spontaneously broken to
\be
G^\nu_{CP} \cong Z^S_2\times H^\nu_{CP} \ ,
\ee
in the neutrino sector. For $G^\nu_{CP}$ to be a well-defined symmetry, the consistency relation of Eq.~\eqref{eq:consistency} needs to be satisfied for the residual symmetry group $Z^S_2$. In other words the elements of $H^\nu_{CP}$ must satisfy,
\begin{equation}\label{eq:z2cons}
X_{\mathbf{r}}\rho^{*}_{\mathbf{r}}(S)X^{-1}_{\mathbf{r}}=\rho_{\mathbf{r}}(S).
\end{equation}
One can arrive at the above restricted consistency equation by recalling the
general consistency condition of Eq.~(\ref{eq:consistency}), with ${g,}g' \in Z^S_2$, and realising that $g'$ must be $S$, as it is the only element which (trivially) has the same order as $g=S$. Armed with this information, recall from section~\ref{sec:2} that a faithful representation of the group $G_f$ must be used to uniquely determine the mapping of the consistency equation (then the unfaithful representations will follow).  Hence, we proceed by considering faithful representations of~$S_4$, i.e. the three dimensional irreducible representations.

For the triplets, one can easily show that for $H^\nu_{CP}$ there are only eight choices for the  $X_{\bf r}$ in Eq.~(\ref{eq:z2cons}) which are acceptable:
\be
X_{\bf r} =\rho_{\bf r}(h) \ ,
\ee
with
\be\label{choicesS4}
h \in  \{1~,~S~,~TST^2~,~T^2ST~,~U~,~SU~,~TST^2U~,~T^2STU\} \ .
\ee
This may be compared to the 24 choices for $X_{\bf r}$ in Eq.~(\ref{24}) corresponding to the original $H_{CP}$ before it is broken (in the charged
lepton sector). Note that the condition in Eq.~(\ref{eq:z2cons}) is automatically satisfied for the doublet and the singlet representations (the unfaithful representations), as is implied by the consistency equation. Comparing the results of Eq.~(\ref{choicesS4}) to the six solutions of the analysis given in~\cite{Feruglio:2012cw}, we observe that our third and fourth solutions of Eq.~(\ref{choicesS4}) are missing in~\cite{Feruglio:2012cw}. This is because the authors of~\cite{Feruglio:2012cw} have made the assumption that a generalised CP transformation must be symmetric. However, it is not necessary to demand symmetric $X_{\bf r}$ if there is a family symmetry, so we keep all eight solutions in the following.

Let us assume that the three generations of lepton doublets reside in the
$\mathbf{3}$-dimensional irreducible representation.\footnote{The results would be the same if the lepton doublets were assigned to $\mathbf{3}'$, since the representation $\mathbf{3}'$ differs from $\mathbf{3}$ only in the overall sign of the generator $U$.} Then, it is possible to construct the
light neutrino mass matrix $m_{\nu}$ by demanding that it respects both the residual flavour symmetry $Z^{S}_2$ and the generalised CP symmetry
$H^\nu_{CP}$:
\begin{eqnarray}
\label{eq:S_falvor}&&\rho^{T}_{\mathbf{3}}(S)m_{\nu}\rho_{\mathbf{3}}(S)=m_{\nu},\\ \label{eq:residual_CP}&&X^{T}_{\mathbf{3}}m_{\nu}X_{\mathbf{3}}=m^{*}_{\nu}.
\end{eqnarray}
The most general matrix which satisfies Eq.~\eqref{eq:S_falvor} has the form
\begin{equation}
\label{eq:S_neutrino}m_\nu ~=~ \alpha \begin{pmatrix} 2&-1&-1\\-1&2&-1\\-1&-1&2\end{pmatrix}
+\beta  \begin{pmatrix}1&0&0\\0&0&1\\0&1&0 \end{pmatrix}
+\gamma \begin{pmatrix}0&1&1\\1&1&0\\1&0&1 \end{pmatrix}
+\epsilon  \begin{pmatrix} 0&1&-1\\1&-1&0\\-1&0&1\end{pmatrix} \ ,
\end{equation}
where $\alpha,\beta,\gamma$, and $\epsilon$ are complex parameters. The CP symmetry of Eq.~\eqref{eq:residual_CP} will further constrain them to be either real or imaginary. The  eight possibilities listed in Eq.~\eqref{choicesS4} are classified in Table~\ref{tab:CP_constraint} according to their four distinct predictions for $\alpha,\beta,\gamma$, and $\epsilon$ being real or imaginary.

\begin{table}[t!]
\begin{center}
\begin{tabular}{|c|c|c|c|c|c|}\hline\hline
  &  $H^{\nu}_{CP}$   &   $\alpha$   &   $\beta$   &   $\gamma$   &   $\epsilon$  \\\hline

(i)  &  $\rho_{\mathbf{r}}(1)$, $\rho_{\mathbf{r}}(S)$  & real  &  real  &  real  &  real  \\ \hline

(ii)  &   $\rho_{\mathbf{r}}(TST^2)$, $\rho_{\mathbf{r}}(T^2ST)$ & imaginary  &  real  &  real  &  real\\  \hline

(iii)  & $\rho_{\mathbf{r}}(U)$, $\rho_{\mathbf{r}}(SU)$  &  real  &  real  &  real  &  imaginary  \\\hline

(iv)  & $\rho_{\mathbf{r}}(TST^2U)$, $\rho_{\mathbf{r}}(T^2STU)$  &  imaginary  &  real  &  real  &  imaginary

\\\hline\hline
\end{tabular}
\caption{\label{tab:CP_constraint}The generalised CP transformations consistent with a residual $Z^{S}_2$ symmetry in the neutrino sector and their corresponding constraints on $\alpha$, $\beta$, $\gamma$ and $\epsilon$ parameters in Eq.~\eqref{eq:S_neutrino}.}
\end{center}
\end{table}

In order to diagonalise $m_{\nu}$ of Eq.~(\ref{eq:S_neutrino}), we perform a tri-bimaximal transformation $U_{TB}$,
\begin{equation}
m'_{\nu}=U^{T}_{TB}m_{\nu}U_{TB}=\left(
\begin{array}{ccc}
 3 \alpha +\beta -\gamma & 0 &
   -\sqrt{3}\epsilon
   \\
 0 & \beta +2\gamma   & 0 \\
 -\sqrt{3}\epsilon &
   0 & 3\alpha-\beta+\gamma
\end{array}
\right),
\end{equation}
with
\begin{equation}
U_{TB}=\left(
\begin{array}{ccc}
 \sqrt{\frac{2}{3}} & \frac{1}{\sqrt{3}}
   & 0 \\
 -\frac{1}{\sqrt{6}} & \frac{1}{\sqrt{3}}
   & -\frac{1}{\sqrt{2}} \\
 -\frac{1}{\sqrt{6}} & \frac{1}{\sqrt{3}}
   & \frac{1}{\sqrt{2}}
\end{array}
\right),
\end{equation}
followed by a (1,3) rotation $U^{\prime}_{\nu}$,
\begin{equation}
U^{\prime T}_{\nu}m^{\prime}_{\nu}U^{\prime}_{\nu}=\text{diag}(m_1,m_2,m_3).
\end{equation}
Then the PMNS matrix is
\be
U_{PMNS}=U_{TB}U^{\prime}_{\nu} \ ,
\ee
where $U_{PMNS}$ is only determined up to the permutations of row and column, since the order of both the charged lepton and neutrino masses cannot be predicted in the present symmetry guided approach, see also \cite{Feruglio:2012cw}. Finally, we shall work in the PDG convention~\cite{pdg},
\begin{equation}
U_{PMNS}=V\,\text{diag}(1,e^{i\frac{\alpha_{21}}{2}},e^{i\frac{\alpha_{31}}{2}}),
\label{eq:pmns_pdg}
\end{equation}
with
\begin{equation}
V=\left(\begin{array}{ccc}
c_{12}c_{13}  &   s_{12}c_{13}   &   s_{13}e^{-i\delta_{CP}}  \\
-s_{12}c_{23}-c_{12}s_{23}s_{13}e^{i\delta_{CP}}   &  c_{12}c_{23}-s_{12}s_{23}s_{13}e^{i\delta_{CP}}  &  s_{23}c_{13}  \\
s_{12}s_{23}-c_{12}c_{23}s_{13}e^{i\delta_{CP}}   & -c_{12}s_{23}-s_{12}c_{23}s_{13}e^{i\delta_{CP}}  &  c_{23}c_{13}
\end{array}\right).
\end{equation}
We are now prepared to consider each of the four cases of
Table~\ref{tab:CP_constraint} in turn.

\begin{description}
\item[~~(i)] $H_{CP}^{\nu}=\rho_{\mathbf{r}}(1)$, $\rho_{\mathbf{r}}(S)$

In this case, the unitary matrix $U^{\prime}_{\nu}$ is
\begin{equation}
\label{eq:i_unup}U^{\prime}_{\nu}=\left(\begin{array}{ccc}
\cos\theta & 0 & \sin\theta  \\
0  & 1 & 0  \\
-\sin\theta &0  &  \cos\theta
\end{array}
\right)P,
\end{equation}
with
\begin{equation}
\tan2\theta=\frac{ \sqrt{3}\;\epsilon}{\beta-\gamma}.\label{theta13theta1}
\end{equation}
The transformation $P$ in Eq. \eqref{eq:i_unup} is a unitary diagonal matrix with entries $\pm1$ or $\pm i$, which renders the light neutrino masses $m_{1,2,3}$ positive. In the following cases, we shall omit this trivial factor. Given the diagonal charged lepton mass matrix, the leptonic mixing matrix is
\begin{equation}
\label{eq:i_pmns}U_{PMNS}=U_{TB}U^{\prime}_{\nu}=\left(
\begin{array}{ccc}
 \sqrt{\frac{2}{3}} \cos\theta &
   \frac{1}{\sqrt{3}} &
   \sqrt{\frac{2}{3}} \sin\theta\\
-\frac{1}{\sqrt{6}}\cos\theta-\frac{1}{\sqrt{2}}\sin\theta & \frac{1}{\sqrt{3}} &
   -\frac{1}{\sqrt{6}}\sin\theta+\frac{1}{\sqrt{2}}\cos\theta \\
 -\frac{1}{\sqrt{6}}\cos\theta+\frac{1}{\sqrt{2}}\sin\theta & \frac{1}{\sqrt{3}} &
   -\frac{1}{\sqrt{6}}\sin\theta -\frac{1}{\sqrt{2}}\cos\theta
 \end{array}
\right){P}\ .
\end{equation}
As a result, the lepton mixing angles and CP phases take the values
\begin{eqnarray}
&\sin\alpha_{21}=\sin\alpha_{31}=\sin\delta_{CP}=0\ , \label{eq:i_angles-ph}\\
\label{eq:i_angles}&\sin^2\theta_{13}=\frac{2}{3}\sin^2\theta,\quad \sin^2\theta_{12}=\frac{1}{2+\cos2\theta},\quad \sin^2\theta_{23}=\frac{1}{2}\left[1+\frac{\sqrt{3}\sin2\theta}{2+\cos2\theta}\right].
\end{eqnarray}
Notice that there is no CP violation in this case, but it is otherwise a viable scenario. Finally, the light neutrino masses are
\begin{eqnarray}
\nonumber&&m_1=\left|3\alpha-\text{sign}\left(\left(\gamma-\beta\right)\cos2\theta\right)\sqrt{(\gamma-\beta)^2+3\epsilon^2}\right| \ ,\\
\nonumber&&m_2=\left|\beta+2\gamma\right|\ , \\
\label{eq:i_masses}&&m_3=\left|3\alpha+\text{sign}\left(\left(\gamma-\beta\right)\cos2\theta\right)\sqrt{(\gamma-\beta)^2+3\epsilon^2}\right|\ .
\end{eqnarray}
The measurement of the reactor mixing angle $\theta_{13}\approx 9^\circ$ fixes the parameter $\theta$ in Eq.~\eqref{eq:i_angles} at
$\theta\approx \pm 11^\circ$.  With this value, the other two mixing angles can be calculated from Eq.~\eqref{eq:i_angles}, yielding $\theta_{12}\approx35.8^\circ$ and $\theta_{23}\approx 45^\circ\pm 6.4^\circ$ which are compatible with the experimentally allowed regions. Concerning the neutrino masses we note that, using Eq.~\eqref{theta13theta1} with $\theta\approx \pm 11^\circ$, it is possible to recast Eq.~\eqref{eq:i_masses} as,
\begin{eqnarray}
\nonumber&&m_1\approx \Big|3\alpha- 1.08 \,(\gamma-\beta) \Big| \ ,\\
\nonumber&&m_2\approx \left|\beta+2\gamma\right|\ , \\
\label{eq:i_masses-replaced}&&m_3\approx
\Big|3\alpha+ 1.08 \,(\gamma-\beta) \Big|
\ .
\end{eqnarray}
As the three neutrino masses are linearly independent combinations of the real parameters $\alpha$, $\beta$ and $\gamma$, any neutrino mass spectrum can be realised in this scenario. Therefore the absolute mass scale as well as the scale of the effective mass relevant in neutrinoless double beta decay cannot be predicted.

\item[~~(ii)] $H_{CP}^{\nu}=\rho_{\mathbf{r}}(TST^2)$, $\rho_{\mathbf{r}}(T^2ST)$

This case is not discussed in~\cite{Feruglio:2012cw} because the authors
require that the CP transformation be both unitary and symmetric. However, it turns out that for this case
\begin{eqnarray*}
m'_{\nu}{m'_{\nu}}^{\dagger}=\left(
\begin{array}{ccc}
 -9 \alpha ^2+(\beta-\gamma)^2+3 \epsilon ^2 & 0 & 0 \\
 0 & (\beta +2 \gamma )^2 & 0 \\
 0 & 0 & -9 \alpha ^2+(\beta-\gamma)^2+3 \epsilon ^2
\end{array}
\right),
\end{eqnarray*}
which implies that $|m_1|^2=|m_3|^2$. Therefore, the resulting light neutrino mass spectrum is (partially) degenerate, and the PMNS matrix cannot be determined uniquely. Phenomenologically, this case is clearly not viable.

\item[~~(iii)] $H_{CP}^{\nu}=\rho_{\mathbf{r}}(U)$, $\rho_{\mathbf{r}}(SU)$

The unitary transformation $U^{\prime}_{\nu}$ is of the form
\begin{equation}
U^{\prime}_{\nu}=\left(
\begin{array}{ccc}
 \cos \theta & 0 & \sin \theta \\
 0 & 1 & 0 \\
 -i\sin \theta & 0 & i\cos \theta
\end{array}
\right),
\end{equation}
with the angle $\theta$
\begin{equation}
\tan2\theta=\frac{i\epsilon}{\sqrt{3}\;\alpha}\ .\label{theta13theta2}
\end{equation}
The PMNS matrix takes the form
\begin{equation}
\label{eq:iii_pmns}U_{PMNS}=U_{TB}U^{\prime}_{\nu}=\left(
\begin{array}{ccc}
 \sqrt{\frac{2}{3}} \cos \theta &  \frac{1}{\sqrt{3}} & \sqrt{\frac{2}{3}} \sin \theta \\
 -\frac{1}{\sqrt{6}}\cos\theta+\frac{i}{\sqrt{2}}\sin\theta & \frac{1}{\sqrt{3}} & -\frac{1}{\sqrt{6}}\sin\theta-\frac{i}{\sqrt{2}}\cos\theta \\
 -\frac{1}{\sqrt{6}}\cos\theta-\frac{i}{\sqrt{2}}\sin \theta &
   \frac{1}{\sqrt{3}} & -\frac{1}{\sqrt{6}}\sin\theta+\frac{i}{\sqrt{2}}\cos\theta
\end{array}
\right).
\end{equation}
Therefore the lepton mixing parameters are
\begin{eqnarray}
&\sin\alpha_{21}=0,\quad \sin\alpha_{31}=0,\quad \left|\sin\delta_{CP}\right|=1,\label{eq:iii_anglesphases}\\
\label{eq:iii_angles}&\sin^2\theta_{13}=\frac{2}{3}\sin^2\theta,\quad \sin^2\theta_{12}=\frac{1}{2+\cos2\theta}, \quad \sin^2\theta_{23}=\frac{1}{2}.
\end{eqnarray}
Notice that we have maximal Dirac CP violation
$\delta_{CP}=\pm\frac{\pi}{2}$. However, the sign cannot be fixed uniquely
because it depends on the value of the angle $\theta$ and the order of the
light neutrino masses. This case provides both reactor and solar mixing angles identical to the ones obtained in case (i) as well as a maximal atmospheric mixing angle, with the light neutrino masses given by
\begin{eqnarray}
\nonumber&&m_1=\left|\beta-\gamma+\text{sign}\left(\alpha\cos2\theta\right)\sqrt{9\alpha^2-3\epsilon^2}\right|\ ,\\
\nonumber&&m_2=\left|\beta+2\gamma\right|\ ,\\
\label{eq:iii_masses}&&m_3=\left|\beta-\gamma-\text{sign}\left(\alpha\cos2\theta\right)\sqrt{9\alpha^2-3\epsilon^2}\right|\ .
\end{eqnarray}
Similar to case (i), the parameter $\epsilon$ in Eq.~\eqref{eq:iii_masses} can be eliminated using Eq.~\eqref{theta13theta2}. The resulting expressions for the three neutrino masses are again linearly independent combinations of the three parameters $\alpha$, $\beta$ and $\gamma$. Therefore, any neutrino masses can be accommodated, and this scenario is viable as well.

\item[~~(iv)] $H_{CP}^{\nu}=\rho_{\mathbf{r}}(TST^2U)$, $\rho_{\mathbf{r}}(T^2STU)$

In this case, the unitary matrix $U^{\prime}_{\nu}$ is given by
\begin{equation}
U^{\prime}_{\nu}=\frac{1}{\sqrt{2}}\left(\begin{array}{ccc}
-e^{i(\theta+\frac{\pi}{4})}   &  0  &   -e^{i(\theta-\frac{\pi}{4})}  \\
0  &  1  &   0  \\
e^{-i(\theta-\frac{\pi}{4})}  &  0   &    -e^{-i(\theta+\frac{\pi}{4})}
\end{array}\right),
\end{equation}
with
\begin{equation}
\tan2\theta=\frac{-\beta+\gamma}{3i\alpha} \ .
\end{equation}
The resulting PMNS matrix reads
\begin{equation}
\label{eq:iv_pmns}\overline{U}_{PMNS}=U_{TB}U^{\prime}_{\nu}=\left(
\begin{array}{ccc}
 -\frac{1}{\sqrt{3}}e^{i \left(\theta +\frac{\pi }{4}\right)} & \frac{1}{\sqrt{3}}
   & -\frac{1}{\sqrt{3}}e^{i \left(\theta -\frac{\pi }{4}\right)} \\
 \frac{1}{2 \sqrt{3}}e^{i \left(\theta+\frac{\pi }{4}\right)}-\frac{1}{2} e^{-i \left(\theta -\frac{\pi }{4}\right)} & \frac{1}{\sqrt{3}} & \frac{1}{2 \sqrt{3}}e^{i\left(\theta -\frac{\pi }{4}\right)}+\frac{1}{2} e^{-i
   \left(\theta +\frac{\pi }{4}\right)} \\
 \frac{1}{2 \sqrt{3}}e^{i \left(\theta+\frac{\pi }{4}\right)}+\frac{1}{2} e^{-i \left(\theta -\frac{\pi }{4}\right)} & \frac{1}{\sqrt{3}} & \frac{1}{2 \sqrt{3}}e^{i\left(\theta -\frac{\pi }{4}\right)}-\frac{1}{2} e^{-i
   \left(\theta +\frac{\pi }{4}\right)}
\end{array}
\right).
\end{equation}
Agreement with the present experimental data can be achieved if we permute the rows of the above $\overline{U}_{PMNS}$ by \cite{Feruglio:2012cw}
\begin{equation}
\label{eq:iv_pmns_ex}U_{PMNS}^{1st}=\left(\begin{array}{ccc}
0&0&1  \\
1&0&0  \\
0&1&0
\end{array}\right)\overline{U}_{PMNS},\quad \text{or}\quad U_{PMNS}^{2nd}=\left(
\begin{array}{ccc}
 0 & 0 & 1 \\
 0 & 1 & 0 \\
 1 & 0 & 0
\end{array}
\right)\overline{U}_{PMNS}\ ,
\end{equation}
to yield two phenomenologically viable PMNS matrices, $U_{PMNS}^{1st}$ and
$U_{PMNS}^{2nd}$, so named because they differ only in their prediction of the atmospheric mixing angle. $U_{PMNS}^{1st}$ predicts the atmospheric mixing angle to be in the first octant and $U_{PMNS}^{2nd}$ in the second. The three lepton mixing angles are then determined to be\footnote{The following results for the mixing angles and CP phases match those of~\cite{Feruglio:2012cw} after making the replacement $\theta\rightarrow \pi/4-\theta$.}
\begin{eqnarray}
\nonumber&&\sin^2\theta_{13}=\frac{1}{6} \left(2-\sqrt{3}
\cos2\theta\right),\qquad\sin^2\theta_{12}=\frac{2}{4+\sqrt{3}\;\cos 2\theta
}\ , \\
\label{eq:iv_angles}&&\sin^2\theta_{23}^{1st}=\frac{2}{4+\sqrt{3}\;\cos
  2\theta },\;
\quad\text{or}\quad\,\sin^2\theta_{23}^{2nd}=1-\frac{2}{4+\sqrt{3}\;\cos
  2\theta }\ .~~~~~~~~
\end{eqnarray}
Both solutions share the phase predictions,
\begin{eqnarray}
\nonumber&&\left|\sin\alpha{_{21}}\right|=\left|\frac{\sqrt{3}+2\cos2\theta}{2+\sqrt{3} \cos2\theta}\right|,\qquad\left|\sin{\alpha'_{31}}\right|=\left|\frac{4\sqrt{3}\;\sin2\theta}{5-3\cos4\theta}\right|,\;\;\\
\label{eq:iv_phases}&&\left|\sin\delta_{{CP}}\right|=\left|\frac{\sqrt{4-2\sqrt{3}\;\cos2\theta}\left(4+\sqrt{3}\cos2\theta\right)\sin2\theta}{5-3\cos4\theta}\right|
\ .
\end{eqnarray}
where $\alpha'_{31}=\alpha_{31}-2\delta_{CP}$, the parameter $\alpha_{31}'$ has been redefined to include $\delta_{CP}$ which is useful in the context of neutrinoless double beta decay \cite{Branco:2011zb}. Lastly, we find the light neutrino masses for this case to be
\begin{eqnarray}
\nonumber&&
m_1=\left|\sqrt{3}\,i\epsilon+\text{sign}\left(i\alpha\cos2\theta\right)\sqrt{(\beta-\gamma)^2-9\alpha^2}\right|\ ,\\
\nonumber&& m_2=\left|\beta+2\gamma\right|\ ,\\
\label{eq:iv_masses}&&
m_3=\left|\sqrt{3}\,i\epsilon-\text{sign}\left(i\alpha\cos2\theta\right)\sqrt{(\beta-\gamma)^2-9\alpha^2}\right|\ .
\end{eqnarray}
From Eq.~\eqref{eq:iv_angles} we see that the smallest value for the reactor angle is obtained when $\theta=0$. Then, $\theta_{13}=12.2^\circ$,
$\theta_{12}=36.2^\circ$ as well as $\theta_{23}^{1 st}=36.2^\circ$ or $\theta_{23}^{2 st}=53.8^\circ$. Clearly, this scenario, which involves non-trivial CP violating phases, is less attractive as it is inconsistent with the 3$\sigma$ allowed region for the reactor angle. Following analogous arguments as before, the neutrino masses remain unconstrained in this case.
\end{description}

\section{\label{sec:s4model2}An effective $\bs{S_4} \rtimes \bs{H_{CP}}$ model }
\cleqn

\begin{table} [h]
\centering
\begin{tabular}{|c||c|c|c|c|c|c||c|c|c|c|c||c|c|c|c|}
\hline \hline
Field &   $L$    &  $N^c$     &  $e^{c}$     &   $\mu^c$    &    $\tau^c$  &  $H_{u,d}$   &  $\varphi_T$   &   $\eta$  &   $\varphi_S$   &  $\xi$   &  $\phi$  &  $\varphi^{0}_T$  &  $\zeta^0$  &  $\varphi^{0}_S$   &  $\xi^{0}$  \\ \hline

$S_4$  &   $\mathbf{3}$  &  $\mathbf{3}$  &  $\mathbf{1}$  & $\mathbf{1}'$   &  $\mathbf{1}$   & $\mathbf{1}$   &   $\mathbf{3}$   &   $\mathbf{2}$   &  $\mathbf{3}'$  &  $\mathbf{1}$  &  $\mathbf{2}$   &  $\mathbf{3}'$  &  $\mathbf{1}$ &  $\mathbf{3}'$   &   $\mathbf{1}$  \\ \hline

$Z_4$  &    1  &   1    &    $i$    &   $-1$   &    $-i$  &  1   &  $i$   &  $i$   &   1   & 1   &  1    &  $-1$  &  $-1$  &  1   &  1 \\ \hline

$Z_3$  &  $\omega^2$   &  $\omega$  &    $\omega$  &   $\omega$  &   $\omega$  &  1  &   1  &   1   &   $\omega$    &  $\omega$  &   $\omega$   & 1   &   1  &  $\omega$  &  $\omega$ \\  \hline

$U(1)_R$  &  1  &   1  &   1  &   1  &   1  &  0 &    0  &  0   & 0  &   0  & 0  &  2 &   2 &   2 &   2   \\\hline \hline
\end{tabular}
\caption{\label{tab:effective}The particle content and the transformation
  properties under the family symmetry $S_4 \times  Z_4\times Z_3$ and $U(1)_R$.}
\end{table}

In this section, we present a realistic effective model of leptons based on
\be
S_4 \rtimes H_{CP}\ ,
\ee
supplemented by the extra symmetries
\be
Z_4\times Z_3 \times U(1)_R
\ee
in order to control the allowed operators in the model.

The three generations of left-handed (LH) lepton doublets $L$ and the three generations of right-handed (RH) neutrinos $N^c$ are both unified into the $S_4$ triplet representation $\mathbf{3}$, while the RH charged leptons are
assigned to be the singlet representations $\mathbf{1}$ or $\mathbf{1}'$. The complete list of lepton, Higgs, flavon and driving fields as well as their transformation properties under the family symmetry are listed in
Table~\ref{tab:effective}.  Notice that the $Z_3$ symmetry is used to separate the flavons entering the charged lepton sector at leading order (LO) from those of the LO neutrino sector. They are further distinguished by an auxiliary $Z_4$ symmetry.  This $Z_4$ symmetry is also helpful in achieving the charged lepton mass hierarchies. As discussed in the previous section, the $S_4$ family symmetry will be spontaneously broken to generate trimaximal mixing.

Furthermore, we will impose a generalised CP symmetry $H_{CP}$ consistent with $S_4$ as discussed in the previous section.  In addition to forcing the coupling constants in the superpotential to be real, the generalised CP symmetry will also be spontaneously broken in the charged lepton sector. In the neutrino sector, a restricted generalised CP symmetry remains,
\be
G^\nu_{CP} \cong Z^S_2\times H^\nu_{CP} \ ,
\ee
where this residual symmetry results in predictions for $\delta_{CP}$ and in other relations as discussed in section~3. However, unlike in the previous section, the dynamics of the spontaneous breaking of $S_4\rtimes H_{CP}$ will be discussed by studying the vacuum alignment of a suitable flavon potential to which we now turn.

\subsection{Vacuum alignment}

The general driving superpotential invariant under the symmetry of the model is
\begin{eqnarray}
\hskip-0.2in w_d&=&g_1\left(\varphi^{0}_T\left(\varphi_T\varphi_T\right)_{\mathbf{3}'}\right)_{\mathbf{1}}+g_2\left(\varphi^{0}_T\left(\eta\varphi_T\right)_{\mathbf{3}'}\right)_{\mathbf{1}}
+g_3\zeta^{0}\left(\varphi_T\varphi_T\right)_{\mathbf{1}}+g_4\zeta^{0}\left(\eta\eta\right)_{\mathbf{1}}\nonumber
\\
&&
+f_1\left(\varphi^{0}_S\left(\varphi_S\varphi_S\right)_{\mathbf{3}'}\right)_{\mathbf{1}}
+f_2\left(\varphi^{0}_S\left(\phi\varphi_S\right)_{\mathbf{3}'}\right)_{\mathbf{1}}+f_3\left(\varphi^{0}_S\varphi_S\right)_{\mathbf{1}}\xi \nonumber
\\
\label{eq:alignment_effective}&&
+h_1\xi^{0}\left(\varphi_S\varphi_S\right)_{\mathbf{1}}+h_2\xi^{0}\left(\phi\phi\right)_{\mathbf{1}}+h_3\xi^0\xi^2+\ldots\,,
\end{eqnarray}
where $(\ldots)_{\mathbf{r}}$ denotes the contraction of the $S_4$ indices to the representation $\mathbf{r}$. If we require the theory to be invariant
under the generalised CP transformation defined above, then all the couplings $g_i$, $f_i$ and $h_i$ are real parameters. The dots in Eq.~\eqref{eq:alignment_effective} stand for higher dimensional operators which are invariant under the flavour symmetry $S_4\times Z_3\times Z_4$ and linear in the driving fields. Due to the constraint of the auxiliary $Z_3$ and $Z_4$ symmetry, the subleading corrections can be obtained by inserting additional flavon fields $\Phi^{~3}_{\nu}$ in all possible ways into the above LO terms, where $\Phi_{\nu}=\left\{\varphi_{S}, \xi, \phi\right\}$ denotes a flavon of the neutrino sector. Therefore the subleading contributions are suppressed by $\langle\Phi_{\nu} \rangle^3/\Lambda^3$ with respect to $w_d$, and can be neglected. In the SUSY limit, the vacuum configuration is determined by the vanishing of the derivative of the driving superpotential $w_d$ with respect to each component of the driving fields. The vacuum in the charged lepton sector is determined by
\begin{eqnarray}
\nonumber\frac{\partial w_d}{\partial\varphi^{0}_{T_1}
}&=&2g_1(\varphi^2_{T_1}-\varphi_{T_2}\varphi_{T_3})+g_2(\eta_1\varphi_{T_2}-\eta_2\varphi_{T_3})=0
\ ,\\
\nonumber\frac{\partial w_d}{\partial\varphi^{0}_{T_2}
}&=&2g_1(\varphi^2_{T_2}-\varphi_{T_1}\varphi_{T_3})+g_2(\eta_1\varphi_{T_1}-\eta_2\varphi_{T_2})=0
\ ,\\
\nonumber\frac{\partial w_d}{\partial\varphi^{0}_{T_3}
}&=&2g_1(\varphi^2_{T_3}-\varphi_{T_1}\varphi_{T_2})+g_2(\eta_1\varphi_{T_3}-\eta_2\varphi_{T_1})=0
\ ,\\
\frac{\partial w_d}{\partial\zeta^{0}}
&=&g_3(\varphi^2_{T_1}+2\varphi_{T_2}\varphi_{T_3})+2g_4\eta_1\eta_2=0 \ .
\end{eqnarray}
This set of equations admits two solutions, the first one is given by
\begin{equation}
\langle\varphi_T\rangle=\left(\begin{array}{ccc}
1\\
1\\
1
\end{array}\right)v_T,\qquad \langle\eta\rangle=\left(\begin{array}{c}1\\
1
\end{array}\right)v_{\eta},\quad \text{with }~~
v^2_T=-\frac{2g_4}{3g_3}v^2_{\eta}\ ,
\end{equation}
and the second solution reads
\begin{equation}
\label{eq:vacuum_charged}\langle\varphi_T\rangle=\left(\begin{array}{c}
0\\
1\\
0
\end{array}\right)v_T,\qquad \langle\eta\rangle=\left(\begin{array}{c}
0\\
1
\end{array}\right)v_{\eta},\quad \text{with} ~~~v_T=\frac{g_2}{2g_1}v_{\eta}\ .
\end{equation}
Here, we choose the second solution.\footnote{Notice that the first solution
can be eliminated by adding another driving field with the same flavour numbers as $\zeta^0$~\cite{Antusch:2011sx}.} The phase of $v_{\eta}$ can be
absorbed into the lepton fields, therefore we can take $v_{\eta}$ to be
real, then the VEV $v_T$ is real as well. The equations determining the vacuum alignment in the neutrino sector are
\begin{eqnarray}
\nonumber\frac{\partial w_d}{\partial
  \varphi^{0}_{S_1}}&=&2f_1(\varphi^2_{S_1}-\varphi_{S_2}\varphi_{S_3})+f_2(\phi_1\varphi_{S_2}+\phi_2\varphi_{S_3})+f_3\xi\varphi_{S_1}=0
\ ,\\
\nonumber\frac{\partial w_d}{\partial
  \varphi^{0}_{S_2}}&=&2f_1(\varphi^2_{S_2}-\varphi_{S_1}\varphi_{S_3})+f_2(\phi_1\varphi_{S_1}+\phi_2\varphi_{S_2})+f_3\xi\varphi_{S_3}=0
\ ,\\
\nonumber\frac{\partial w_d}{\partial
  \varphi^{0}_{S_3}}&=&2f_1(\varphi^2_{S_3}-\varphi_{S_1}\varphi_{S_2})+f_2(\phi_1\varphi_{S_3}+\phi_2\varphi_{S_1})+f_3\xi\varphi_{S_2}=0
\ ,\\
\frac{\partial
  w_d}{\partial\xi^0}&=&h_1(\varphi^2_{S_1}+2\varphi_{S_2}\varphi_{S_3})+2h_2\phi_1\phi_2+h_3\xi^2=0
\ .
\end{eqnarray}
Disregarding the ambiguity caused by the $S_4$ family symmetry transformations, we find the solution
\begin{equation}
\label{eq:vacuum_neutrino}\langle\varphi_S\rangle=\left(\begin{array}{ccc}
1\\
1\\
1
\end{array}\right)v_S,\qquad \langle\phi\rangle=\left(\begin{array}{c}
v_1\\
v_2
\end{array}\right),\qquad \langle\xi\rangle=u\ ,
\end{equation}
where the VEVs obey the relation
\begin{eqnarray}
\label{eq:plus1}&&v_1+v_2=-f_3u/f_2\equiv -Bu \ ,\\
\label{eq:linear1}&&3h_1v^2_S+2h_2v_1v_2=-h_3u^2 \ .
\end{eqnarray}
The VEVs $v_S$, $v_1$ and $v_2$ are undetermined. In order to solve this problem, we introduce a second driving fields with identical quantum numbers
as $\xi^0$~\cite{Antusch:2011sx}. Consequently we obtain two $F$-term conditions which are identical in their structures but involve independent
coupling constants, $h_i$ for one driving field and $h'_i$ for the other,
i.e.
\begin{equation}
\label{eq:linear2}3h'_1v^2_S+2h'_2v_1v_2=-h'_3u^2\ .
\end{equation}
As both conditions, Eq.~\eqref{eq:linear1} and Eq.~\eqref{eq:linear2}, must be satisfied, one can find a unique solution for the VEV of the $\varphi_S$ flavon
\begin{equation}
v^2_S=\frac{h_2h'_3-h_3h'_2}{3\left(h_1h'_2-h_2h'_1\right)}u^2\ ,
\end{equation}
and
\begin{equation}
\label{eq:multiply1}v_1v_2=\frac{h_3h'_1-h_1h'_3}{2\left(h_1h'_2-h_2h'_1\right)}u^2\equiv
Cu^2\ .
\end{equation}
Since it is always possible to absorb the phase of $u$ by a redefinition of
the matter fields, we can take $u$ to be real without loss of generality, then $v_S$ is either real or purely imaginary. The vacuum of the doublet $\phi$ can be obtained by solving Eq.~(\ref{eq:plus1}) together with Eq.~(\ref{eq:multiply1}).
\begin{itemize}
  \item {For $B^2-4C>0$ we find}

  \begin{equation}
  \label{eq:v1_v2_real}\left\{\begin{array}{c}
  v_1=\frac{1}{2}\left[-B\pm\sqrt{B^2-4C}\;\right]u \ ,\\
  v_2=\frac{1}{2}\left[-B\mp\sqrt{B^2-4C}\;\right]u \ ,
  \end{array}\right.
  \end{equation}
from which we see that $v_1$ and $v_2$ are both real.

  \item{For $B^2-4C<0$ we obtain}

  \begin{equation}
  \label{eq:v1_v1_conjugate}\left\{\begin{array}{c}
  v_1=\frac{1}{2}\left[-B\pm i\sqrt{4C-B^2}\;\right]u \ ,\\
  v_2=\frac{1}{2}\left[-B\mp i\sqrt{4C-B^2}\;\right]u \ ,
  \end{array}\right.
  \end{equation}
which entails $v_1=v^{*}_2$ in this case.
\end{itemize}

\subsection{The lepton masses and mixings}

The effective superpotential for the charged lepton masses which is allowed by the symmetries is given as
\begin{eqnarray}
w_{\ell}&=&\frac{y_{\tau}}{\Lambda}\left(L\varphi_T\right)_{\mathbf{1}}H_d\tau^{c}+\frac{y_{\mu_1}}{\Lambda^2}\left(L\left(\varphi_T\varphi_T\right)_{\mathbf{3}'}\right)_{\mathbf{1}'}H_d\mu^{c}
+\frac{y_{\mu_2}}{\Lambda^2}\left(L\left(\eta\varphi_T\right)_{\mathbf{3}'}\right)_{\mathbf{1}'}H_d\mu^{c}\nonumber\\
&&+\sum_{i}\frac{y_{e_i}}{\Lambda^3}\left(L\mathcal{O}_i\right)_{\mathbf{1}}H_de^{c}+\ldots\ ,\label{eq:fivecon}
\end{eqnarray}
where $\mathcal{O}$ stands for
\begin{equation}
\mathcal{O}=\left\{\varphi^3_T,\;\eta\varphi^2_T,\;\eta^2\varphi_T\right\} \ ,
\end{equation}
and all possible $S_4$ contractions are to be considered. The dots in  Eq.~\eqref{eq:fivecon} denote subleading corrections to the Yukawa superpotential $w_{\ell}$ which can be obtained by multiplying all LO terms of Eq.~\eqref{eq:fivecon} by $\Phi^{~3}_{\nu}/\Lambda^3$, with $\Phi_{\nu}=\left\{\varphi_{S}, \xi, \phi\right\}$. The resulting corrections are of relative order $\langle\Phi_\nu\rangle^3/\Lambda^3$ with respect to the LO terms and therefore negligible. It is interesting to note that the tau mass is suppressed by
$1/\Lambda$, while the muon and
electron masses appear at order $1/\Lambda^2$ and $1/\Lambda^3$, respectively. The mass hierarchies among the charged lepton are thus produced in a natural way. Inserting the vacuum alignment of Eq.~(\ref{eq:vacuum_charged}), we find a diagonal charged lepton mass matrix with
\begin{eqnarray}
m_\tau=y_{\tau}\frac{v_T}{\Lambda}v_d,\qquad
m_{\mu}=\left[2y_{\mu_1}\left(\frac{v_T}{\Lambda}\right)^2-y_{\mu_2}\frac{v_{\eta}v_T}{\Lambda^2}\right]v_d,\qquad
m_e=y_e\left(\frac{v_T}{\Lambda}\right)^3v_d,
\end{eqnarray}
where $y_e$ is the result five different contributions corresponding to the $y_{e_i}$ in Eq.~\eqref{eq:fivecon}. Obviously the VEVs of the flavons $\varphi_T$ and $\eta$ are responsible for the spontaneous breaking of  both the family symmetry and  the generalised CP symmetry. It is straightforward to check that $S_4$ is broken completely in the charged lepton sector. In addition, only one CP symmetry $X_{\mathbf{r}}=\rho_{\mathbf{r}}(1)$ out of the 24 consistent CP transformations is preserved by the flavons $\varphi_T$ and $\eta$, given the previously mentioned fact that both $v_{T}$ and $v_{\eta}$ can be chosen to be real. As a consequence, the residual CP symmetry in the charged lepton sector is $H_{CP}^{\ell}=\rho_{\mathbf{r}}(1)$. Now we turn to the neutrino sector; the LO effective superpotential is given by
\begin{equation}
w_{\nu}~=~y\left(LN^c\right)_{\mathbf{1}}H_u+y_1\left(\left(N^cN^c\right)_{\mathbf{3}'}\varphi_S\right)_{\mathbf{1}}+y_2\left(N^cN^c\right)_{\mathbf{1}}\xi+y_3\left(\left(N^cN^c\right)_{\mathbf{2}}\phi\right)_{\mathbf{1}}
\ ,
\end{equation}
where all the couplings $y_i$ are real due to the general CP invariance. The subleading operators contributing to the Dirac and the right-handed Majorana neutrino masses are of the form $\left(LN^c\Phi^{~3}_{\nu}\right)_{\mathbf{1}}H_u/\Lambda^3$ and $\left(N^cN^c\Phi^{~4}_{\nu}\right)_{\mathbf{1}}/\Lambda^4$, respectively. Hence the higher order corrections can again be safely neglected. At LO, the Dirac neutrino mass matrix takes a trivial form
\begin{equation}
m_D=\left(\begin{array}{ccc}
1&0&0  \\
0&0&1 \\
0&1&0
\end{array}\right)yv_u \ .
\end{equation}
Given the vacuum configuration of Eq.~(\ref{eq:vacuum_neutrino}), it is
straightforward to derive the RH neutrino mass matrix
\begin{equation}
m_M=a\left(
\begin{array}{ccc}
 2 & -1 & -1 \\
 -1 & 2 & -1 \\
 -1 & -1 & 2
\end{array}
\right)+b \left(
\begin{array}{ccc}
 1 & 0 & 0 \\
 0 & 0 & 1 \\
 0 & 1 & 0
\end{array}
\right)+c\left(
\begin{array}{ccc}
 0 & 0 & 1 \\
 0 & 1 & 0 \\
 1 & 0 & 0
\end{array}
\right)+d \left(
\begin{array}{ccc}
 0 & 1 & 0 \\
 1 & 0 & 0 \\
 0 & 0 & 1
\end{array}
\right),
\end{equation}
where we have introduced the parameters
\begin{equation}
a=y_1v_S,\qquad b=y_2u,\qquad c=y_3v_2,\qquad d=y_3v_1 \ .
\end{equation}
The light neutrino mass matrix is then obtained from the seesaw formula, yielding
\begin{eqnarray}
\nonumber\hskip-0.3in m_{\nu}&=&-m_Dm^{-1}_Mm^{T}_D\\
\label{eq:neutrino_mass_effective}\hskip-0.2in&=&\alpha \begin{pmatrix} 2&-1&-1\\-1&2&-1\\-1&-1&2\end{pmatrix}
+\beta  \begin{pmatrix}1&0&0\\0&0&1\\0&1&0 \end{pmatrix}
+\gamma \begin{pmatrix}0&1&1\\1&1&0\\1&0&1 \end{pmatrix}
+\epsilon  \begin{pmatrix} 0&1&-1\\1&-1&0\\-1&0&1\end{pmatrix}.~~~
\end{eqnarray}
This matrix is of the same form as the neutrino mass matrix in Eq.~\eqref{eq:S_neutrino} which is the most general matrix invariant under the action of the $S_4$ element $S$. The parameters $\alpha$, $\beta$, $\gamma$ and $\epsilon$ are related to $a$, $b$, $c$ and $d$ by
\begin{eqnarray}
\nonumber\alpha&=&\frac{a}{-9 a^2+b^2+(c+d)^2-3cd-b(c+d)}\,,\\
\nonumber\beta&=&\frac{3 a^2-b^2+c d}{(b+c+d) \left[-9 a^2+b^2+(c+d)^2-3cd-b(c+d)\right]}\,,\\
\nonumber\gamma&=&\frac{6 a^2-(c+d)^2+2cd+b(c+d)}{2 (b+c+d) \left[-9 a^2+b^2+(c+d)^2-3cd-b(c+d) \right]}\,,\\
\label{eq:a_alpha}\epsilon&=&\frac{d-c}{2 \left[-9
    a^2+b^2+(c+d)^2-3cd-b(c+d)\right]}\ ,
\end{eqnarray}
where the overall factor $y^2v^2_u$ in these expressions has been omitted. From the discussion of the vacuum alignment, we know that $b$ is a real parameter, and $a$ can be real or purely imaginary. For the doublet flavon $\phi$, the VEVs $v_1$ and $v_2$ can be both real for the first solution in
Eq.~(\ref{eq:v1_v2_real}), yielding real $c$ and $d$, or, for the second
solution shown in Eq.~(\ref{eq:v1_v1_conjugate}), $v_1$ and $v_2$ are complex conjugates of each other, yielding $c=d^{*}$. Therefore there are only four possible cases allowed in our model, which are listed in Table \ref{tab:effective_model}.

\begin{table}[t!]
\begin{center}
{\footnotesize
\begin{tabular}{|c|c|c|c|c|c|}\hline\hline
&   &  $\alpha$ &   $\beta$   &   $\gamma$   &   $\epsilon$     \\ \hline

(i) & $a\in\mathbb{R},\,b\in\mathbb{R},\,c\in\mathbb{R},\,d\in\mathbb{R}$  &  real &  real &   real   &  real  \\ \hline

(ii) & $a\in\mathbb{C},\,b\in\mathbb{R},\,c\in\mathbb{R},\,d\in\mathbb{R}$ \text{with} $a=-a^{*}$  &  imaginary  &   real   &  real  &  real   \\ \hline

(iii) & $a\in\mathbb{R},\,b\in\mathbb{R},\,c\in\mathbb{C},\,d\in\mathbb{C}$ \text{with} $c=d^{*}$  & real  &   real   &  real &  imaginary   \\  \hline

(iv) & $a\in\mathbb{C},\,b\in\mathbb{R},\,c\in\mathbb{C},\,d\in\mathbb{C}$ \text{with} $a=-a^{*}$ and $c=d^{*}$  &  imaginary  &  real   &  real  & imaginary  \\ \hline \hline
\end{tabular}
}	
\caption{\label{tab:effective_model}The four possible cases of the effective model which are controlled by the input parameters
of the flavon potential.}
\end{center}
\end{table}

In order to understand the phenomenological implications of the model, it is useful to know how the generalised CP symmetry is spontaneously broken and what the remnant CP symmetry is. Once the flavon fields acquire their VEVs, the $S_4$ family symmetry and the generalised CP symmetry are spontaneously broken. Imagining that the flavon VEVs could transform in the way as the flavon fields under the action of CP transformation, i.e.
\begin{eqnarray}
\nonumber&&\langle\varphi_S\rangle=\left(\begin{array}{c}
1\\
1\\
1
\end{array}\right)v_S\stackrel{CP}{\longrightarrow}\rho_{\mathbf{3}'}(g)\langle\varphi_S\rangle^{*}=\rho_{\mathbf{3}'}(g)\left(\begin{array}{c}
1\\
1\\
1
\end{array}\right)v^{*}_S\;,\quad g\in S_4\ ,\\
&&\langle\phi\rangle=\left(\begin{array}{c}
v_1 \\
v_2
\end{array}\right)\stackrel{CP}{\longrightarrow}\rho_{\mathbf{2}}(g)\langle\phi\rangle^{*}=\rho_{\mathbf{2}}(g)\left(\begin{array}{c}
v^{*}_1 \\
v^{*}_2
\end{array}\right),\qquad \;
\langle\xi\rangle=u\stackrel{CP}{\longrightarrow}\langle\xi\rangle^{*}=u^{*}\ ,~~~~~~~~~~
\end{eqnarray}
then all the initial 24 consistent CP symmetries would be kept. However, the flavon VEVs are only numbers and they don't change at all under a CP transformation. Therefore only those CP transformations which transform the
corresponding flavon VEVs into themselves remain symmetries of the theory after symmetry breaking. That means, the residual CP symmetry $\rho_{\mathbf{r}}(g)$ in the neutrino sector satisfies
\begin{equation}
\rho_{\mathbf{3}'}(g)\langle\varphi_S\rangle^{*}=\langle\varphi_S\rangle,\quad
\rho_{\mathbf{2}}(g)\langle\phi\rangle^{*}=\langle\phi\rangle,\quad
\rho_{\mathbf{1}}(g)\langle\xi\rangle^{*}=\langle\xi\rangle^{*}=\langle\xi\rangle\ .
\end{equation}
In the following, we discuss the four cases in Table~\ref{tab:effective_model} one by one.

\begin{description}
\item[~~(i)]
In this case, both the triplet VEV $v_S $ and the doublet VEVs $v_1$ and $v_2$ are real. We can easily check that the generalised CP symmetries for
$h=1$ and $h=S$ are preserved in the neutrino sector, i.e. the residual CP
symmetry in the neutrino sector is $H_{CP}^{\nu}=\left\{\rho_{\mathbf{r}}({1}),\,\rho_{\mathbf{r}}(S)\right\}$. From Eq.~\eqref{eq:a_alpha} we see that all four parameters $\alpha$, $\beta$, $\gamma$ and $\epsilon$ are real. This is exactly the case (i) discussed in the general analysis of section~3 which is solely based on symmetry arguments. As a result, the PMNS matrix and the resulting lepton mixing parameters are of the form shown in Eqs.~(\ref{eq:i_pmns}-\ref{eq:i_angles}). In the present case, the generalised CP symmetry $X_{\mathbf{r}}=\rho_{\mathbf{r}}(1)$ is preserved in both the neutrino and the charged lepton sector. According to the general results for weak basis invariants stated in Appendix \ref{app1}, the CP phases would be trivial, i.e. although this case is viable, there is no CP violation, as has been shown already in Eq.~\eqref{eq:i_angles-ph}.

\item[~~(ii)]
In this case, $v_S$ is purely imaginary and the remaining VEVs $v_1$ and $v_2$ are real. The generalised CP symmetry is broken to $H_{CP}^{\nu}=\left\{\rho_{\mathbf{r}}(TST^2),\;\rho_{\mathbf{r}}(T^2ST)\right\}$
in the neutrino sector. Concerning the light neutrino mass matrix, the parameter $\alpha$ is imaginary, while $\beta$, $\gamma$ and $\epsilon$ are
all real. This is exactly case (ii) of the general analysis of section~3, where the light neutrino masses are degenerate and hence this case is not viable.

\item[~~(iii)]
This case corresponds to the VEV $v_S$ being real together with the solution $v_1=v^{*}_2$ for the doublet $\phi$. Now only two of the 24 generalised CP symmetries are preserved in the neutrino sector and $H_{CP}^{\nu}=\left\{\rho_{\mathbf{r}}(U),\;\rho_{\mathbf{r}}(SU)\right\}$. The light neutrino mass matrix is of the form given in Eq.~\eqref{eq:neutrino_mass_effective}, where $\alpha$, $\beta$ and $\gamma$
are real while $\epsilon$ is purely imaginary. This corresponds to case (iii) studied in section~\ref{sec:general}, and thus the predictions for the PMNS matrix in Eq.~\eqref{eq:iii_pmns} with lepton mixing parameters as in
Eqs.~(\ref{eq:iii_anglesphases},\ref{eq:iii_angles}). The neutrino mixing is
of trimaximal form, and it is remarkable that we obtain maximal Dirac CP violation $\delta_{CP}=\pm \pi/2$ in this case. As for the generalised CP symmetry breaking, although the CP symmetries $H_{CP}^{\ell}=\left\{\rho_{\mathbf{r}}(1)\right\}$ and
$H_{CP}^{\nu}=\left\{\rho_{\mathbf{r}}(U),\;\rho_{\mathbf{r}}(SU)\right\}$ are preserved in the charged lepton and the neutrino sector, respectively,
the CP symmetry is completely broken in the full theory, and the mismatch
between $H_{CP}^{\ell}$ and $H_{CP}^{\nu}$ is precisely the origin of the
maximal Dirac CP violation. Therefore this case is viable and predicts maximal CP violation.

\item[~~(iv)]
This case corresponds to $v_S$ being purely imaginary combined with the
solution $v_1=v^{*}_2$. The remnant CP symmetry in the neutrino sector becomes $H_{CP}^{\nu}=\left\{\rho_{\mathbf{r}}(TST^2U),\; \rho_{\mathbf{r}}(T^2STU)\right\}$. The parameters $\alpha$ and $\epsilon$ of the corresponding light neutrino mass matrix  are imaginary, while $\beta$ and $\gamma$ are real. This is exactly the case (iv) investigated in the general analysis of section~\ref{sec:general}. As has been pointed out, in order to achieve agreement with the present experimental data, one has to permute the rows of $U_{PMNS}$ as done in Eq.~\eqref{eq:iv_pmns_ex}. This permutation corresponds to exchanging the three charged lepton masses. However, in the present model, the charged lepton masses are predicted to be of different orders in the expansion parameter $v_{T}/\Lambda$ ($v_{\eta}/\Lambda$), so that this permutation is forbidden. Although we can permute its columns, as the neutrino mass spectrum can be either normal or inverted, the resulting PMNS matrix always leads to $\sin^2\theta_{13}=1/3$, which is much larger than the measured value. Hence in the framework of the effective model, this case is not viable.

\end{description}

Finally, we note that the residual family symmetry in the neutrino sector is
$G_{\nu}=Z^{S}_2\equiv\left\{{1},S\right\}$ for all the four cases discussed. This is the reason why the second column of the tri-bimaximal mixing matrix is kept. In summary, in the effective model there are only two
viable cases, namely (i) with no CP violation, and (iii) with maximal CP violation $\delta_{CP}=\pm {\pi}/{2}$.

\section{A renormalisable $\bs{S_4} \rtimes \bs{H_{CP}}$ model}
\cleqn

It is generally believed that the fundamental theory formulated at a high energy scale should be renormalisable. Any non-renormalisable operators of the effective low-energy theory should then arise from the fundamental underlying theory by integrating out the heavy degrees of freedom. Therefore, we present an improved renormalisable model in this section, where tri-bimaximal neutrino mixing is produced at leading order, while next-to-leading order (NLO) contributions break the tri-bimaximal to a trimaximal mixing pattern. This will naturally explain why the reactor angle and the deviations from maximal  atmospheric mixing are relatively small.

The model of this section is inspired by the renormalisable trimaximal $S_4$ model of~\cite{King:2011zj} which was originally proposed without $H_{CP}$. Here we shall construct an analogous renormalisable $S_4$ model but with generalised CP symmetry based on
\be
S_4 \rtimes H_{CP} \ ,
\ee
supplemented by the same extra symmetries as in the previous effective model, i.e.
\be
Z_4\times Z_3 \times U(1)_R\ ,
\ee
in order to control the allowed operators in the model.  The matter fields, flavon fields and their transformation properties under the imposed symmetries are presented in Table \ref{tab:renormalisable}, while the driving fields,  messenger fields, and their transformation properties are given in Table~\ref{tab:messenger}. Tables \ref{tab:renormalisable} and \ref{tab:messenger} may be compared to Table~\ref{tab:effective} of the effective model.

As before we impose a generalised CP symmetry $H_{CP}$ consistently with $S_4$. In addition to forcing the couplings in the superpotential to be real, the generalised CP symmetry will be spontaneously broken.  The remaining symmetry in the neutrino sector,
\be
G^\nu_{CP} \cong Z^S_2\times H^\nu_{CP},
\ee
will result in predictions for $\delta_{CP}$ and  other relations, as discussed in the previous section. However in order to justify this breaking, we must perform a detailed analysis of the flavon potential of the renormalisable model.

\begin{table} [t!]
\centering
\begin{tabular}{|c||c|c|c|c|c|c||c|c|c|c|c|c|}
\hline \hline

Field &   ~~$L$~~    &   ~$N^c$~     &  ~$e^{c}$~     &   ~$\mu^c$~     &    ~$\tau^c$~  &  $H_{u,d}$ &  ~$\varphi_T$~   &  ~~$\eta$~~  &   ~$\varphi_S$~  & ~~$\phi$~~  &  ~~$\xi$~~  &   ~$\Delta$~   \\  \hline

$S_4$  &  $\mathbf{3}$ &  $\mathbf{3}$  &  $\mathbf{1}$    &    $\mathbf{1}'$   &    $\mathbf{1}$  &    $\mathbf{1}$  &  $\mathbf{3}$  &  $\mathbf{2}$   &   $\mathbf{3}'$  &   $\mathbf{2}$  &   $\mathbf{1}$  &  $\mathbf{1}'$ \\  \hline

$Z_4$   &  1         &    1  &  $i$   &  $-1$   &   $-i$  & 1  &   $i$  &  $i$  &  1   &   1  &  1  &   $1$ \\ \hline

$Z_3$  &  $\omega$   &     $\omega^2$   &   $\omega^2$   &  1  &  $\omega$  &  1   &  $\omega$  &   $\omega$  &   $\omega^2$  &   $\omega^2$  &  $\omega^2$    &   1 \\   \hline

$U(1)_R$ &  1  &   1  &   1  &  1   &  1  &  0  &  0  &   0   &   0     &  0 &  0   &   0    \\

\hline \hline

\end{tabular}
\caption{\label{tab:renormalisable}The transformation properties of the matter fields, Higgs and flavon fields in the renormalisable model.}
\end{table}

\subsection{\label{subsec:vacuum_renor}Vacuum alignment }

The renormalisable driving superpotential that is linear in the driving fields and invariant under the flavour symmetry is
\begin{eqnarray}
\nonumber w_d&=&
g_1\left(\varphi^{0}_T\left(\varphi_T\varphi_T\right)_{\mathbf{3}'}\right)_{\mathbf{1}}+
g_2\left(\varphi^{0}_T\left(\eta\varphi_T\right)_{\mathbf{3}'}\right)_{\mathbf{1}}
+g_3\zeta^{0}\left(\varphi_T\varphi_T\right)_{\mathbf{1}}
+g_4\zeta^{0}\left(\eta\eta\right)_{\mathbf{1}}\\
&&+f_1\left(\varphi^{0}_S\left(\varphi_S\varphi_S\right)_{\mathbf{3}'}\right)_{\mathbf{1}}\nonumber
+f_2\left(\varphi^{0}_S\left(\phi\varphi_S\right)_{\mathbf{3}'}\right)_{\mathbf{1}}
+f_3\left(\varphi^{0}_S\varphi_S\right)_{\mathbf{1}}\xi
+f_4\left(\widetilde{\varphi}^{\;0}_S\left(\phi\varphi_S\right)_{\mathbf{3}}\right)_{\mathbf{1}}\\
&&+f_5\xi^{0}\left(\varphi_S\varphi_S\right)_{\mathbf{1}}+f_6\xi^{0}\left(\phi\phi\right)_{\mathbf{1}}+f_7\xi^0\xi^2
\label{eq:alignment_renor}+M^2\Delta^0+f_8\Delta^0\Delta^2 \ ,
\end{eqnarray}
where the term $\Delta^0H_uH_d$ has been neglected, as it will play no role in the flavon vacuum alignment because the breaking of $S_4$ flavour symmetry is typically assumed to occur around the GUT scale. Furthermore, the couplings $g_i$, $f_i$ and $M$ are real parameters due to the generalised CP symmetry.

Similar to the effective model of section \ref{sec:s4model2}, the vacuum alignment for the flavons of the charged lepton sector  is determined by the $F$-term conditions of $\varphi^0_T$ and $\zeta^{0}$:
\begin{eqnarray}
\nonumber \frac{\partial w_d}{\partial\varphi^{0}_{T_1}
}&=&2g_1(\varphi^2_{T_1}-\varphi_{T_2}\varphi_{T_3})+g_2(\eta_1\varphi_{T_2}-\eta_2\varphi_{T_3})=0
\ ,\\
\nonumber \frac{\partial w_d}{\partial\varphi^{0}_{T_2}
}&=&2g_1(\varphi^2_{T_2}-\varphi_{T_1}\varphi_{T_3})+g_2(\eta_1\varphi_{T_1}-\eta_2\varphi_{T_2})=0
\ , \\
\nonumber \frac{\partial w_d}{\partial\varphi^{0}_{T_3}
}&=&2g_1(\varphi^2_{T_3}-\varphi_{T_1}\varphi_{T_2})+g_2(\eta_1\varphi_{T_3}-\eta_2\varphi_{T_1})=0
\ ,\\
\nonumber \frac{\partial w_d}{\partial\zeta^{0}}
&=&g_3(\varphi^2_{T_1}+2\varphi_{T_2}\varphi_{T_3})+2g_4\eta_1\eta_2=0 \ .
\end{eqnarray}
This set of equations is satisfied by two solutions, where the ambiguity caused by $S_4$ symmetry transformations is ignored. The first solution is given by
\begin{equation}
\langle\varphi_T\rangle=\left(\begin{array}{ccc}
1\\
1\\
1
\end{array}\right)v_T,\qquad \langle\eta\rangle=\left(\begin{array}{c}1\\
1
\end{array}\right)v_{\eta},\quad \text{with }~~
v^2_T=-\frac{2g_4}{3g_3}v^2_{\eta}\ ,
\end{equation}
while the second is given by
\begin{equation}
\label{eq:vacuum_charged_renor}\langle\varphi_T\rangle=\left(\begin{array}{c}
0\\
1\\
0
\end{array}\right)v_T,\qquad \langle\eta\rangle=\left(\begin{array}{c}
0\\
1
\end{array}\right)v_{\eta},\quad \text{with} ~~~v_T=\frac{g_2}{2g_1}v_{\eta}
\ .
\end{equation}
In this work, we choose the second solution since the first solution can be removed by introducing another driving field that transforms identically with $\zeta^0$, similar to the effective model. Notice that the phase of $v_{\eta}$ can be absorbed into the lepton fields. Therefore, we take $v_{\eta}$ to be real, implying that the VEV $v_T$ is real as well.

We continue the analysis of the vacuum alignment  by considering the flavon fields associated with the neutrino sector, i.e. $\varphi_S$, $\phi$ and $\xi$. The $F$-term conditions determining their alignments are \begin{eqnarray}
\nonumber\frac{\partial w_d}{\partial
  \varphi^{0}_{S_1}}&=&2f_1(\varphi^2_{S_1}-\varphi_{S_2}\varphi_{S_3})+f_2(\phi_1\varphi_{S_2}+\phi_2\varphi_{S_3})+f_3\xi\varphi_{S_1}=0
\ ,\\
\nonumber\frac{\partial w_d}{\partial \varphi^{0}_{S_2}}&=&2f_1(\varphi^2_{S_2}-\varphi_{S_1}\varphi_{S_3})+f_2(\phi_1\varphi_{S_1}+\phi_2\varphi_{S_2})+f_3\xi\varphi_{S_3}=0 \ ,\\
\nonumber\frac{\partial w_d}{\partial \varphi^{0}_{S_3}}&=&2f_1(\varphi^2_{S_3}-\varphi_{S_1}\varphi_{S_2})+f_2(\phi_1\varphi_{S_3}+\phi_2\varphi_{S_1})+f_3\xi\varphi_{S_2}=0\ , \\
\nonumber\frac{\partial w_d}{\partial \widetilde{\varphi}^{0}_{S_1}}&=&f_4\left(\phi_1\varphi_{S_2}-\phi_2\varphi_{S_3}\right)=0 \ , \\
\nonumber\frac{\partial w_d}{\partial \widetilde{\varphi}^{0}_{S_2}}&=&f_4\left(\phi_1\varphi_{S_1}-\phi_2\varphi_{S_2}\right)=0 \ , \\
\nonumber\frac{\partial w_d}{\partial \widetilde{\varphi}^{0}_{S_3}}&=&f_4\left(\phi_1\varphi_{S_3}-\phi_2\varphi_{S_1}\right)=0 \ , \\
\frac{\partial
  w_d}{\partial\xi^0}&=&f_5(\varphi^2_{S_1}+2\varphi_{S_2}\varphi_{S_3})+2f_6\phi_1\phi_2+f_7\xi^2=0\,.
\end{eqnarray}
This set of equations leads to vacuum alignments given by
\begin{equation}
\label{eq:vacuum_neu_renor}\langle\varphi_S\rangle=\left(\begin{array}{c}
1\\
1\\
1
\end{array}\right)v_S,\qquad \langle\phi\rangle=\left(\begin{array}{c}
1\\
1
\end{array}\right)v_{\phi},\qquad   \langle\xi\rangle=u\ ,
\end{equation}
where
\begin{equation}
\label{eq:vev_relation}v^2_S=-\frac{1}{6f^2_2f_5}\left(f^2_3f_6+2f^2_2f_7\right)u^2,\qquad
v_{\phi}=-\frac{f_3}{2f_2}u\ .
\end{equation}
Notice that the vacuum alignments of the flavons  shown in Eq.~(\ref{eq:vacuum_neu_renor}) are invariant under the action of both the $S$ and $U$ elements of $S_4$, preserving the tri-bimaximal Klein four subgroup. Furthermore, the phase of $u$ can be absorbed by field redefinition. Hence, $u$ can be taken to be real without loss of generality. This renders $v_{\phi}$  real as well, but the VEV $v_S$ can be real or purely imaginary depending on the coefficient $-\left(f^2_3f_6+2f^2_2f_7\right)/(f^2_2f_5)$ being positive or negative, respectively.

We conclude the discussion of the vacuum alignment of the renormalisable model by considering the last two operators of Eq.~(\ref{eq:alignment_renor}) which are responsible for the alignment of the flavon $\Delta$. They provide the $F$-term minimisation condition
\begin{eqnarray}
\frac{\partial w_d}{\partial\Delta^0 }=M^2+f_8\Delta^2=0\ ,
\end{eqnarray}
which is satisfied when
\begin{equation}
v^2_{\Delta}=-M^2/f_8~,
\end{equation}
where $v_{\Delta}=\langle\Delta\rangle$. Notice that if $f_8>0$, then $v_{\Delta}$ is purely imaginary. However if $f_8<0$, then $v_{\Delta}$ is completely real. Hence, $v_{\Delta}$ is constrained to be either real or purely imaginary in this model. Thus, having completed the discussion of the vacuum alignment of the renormalisable model, we now proceed to investigate the leptonic masses and mixings predicted by it.

\begin{table}[t!]
\centering
\begin{tabular}{|c||c|c|c|c|c|c||c|c|c|c|c|c|c|c|}
\hline \hline

Field &  $\varphi^{0}_T$   &  $\zeta^0$   &   $\varphi^0_S$  &   $\widetilde{\varphi}^{\,0}_S$   &  $\xi^0$  &    $\Delta^0$  &   $\Omega_1$   &   $\Omega^{c}_1$   &   $\Omega_2$   &   $\Omega^c_2$  &  $\Omega_3$   &   $\Omega^{c}_3$  &   $\Sigma$   &   $\Sigma^c$   \\  \hline

$S_4$   &    $\mathbf{3}'$   &    $\mathbf{1}$   &   $\mathbf{3}'$   &   $\mathbf{3}$  &   $\mathbf{1}$  &   $\mathbf{1}$  &   $\mathbf{2}$  &   $\mathbf{2}$ &  $\mathbf{2}$   &  $\mathbf{2}$ &  $\mathbf{3}$  &   $\mathbf{3}$   &  $\mathbf{3}'$   &  $\mathbf{3}'$   \\ \hline

$Z_4$   &   $-1$     &      $-1$   & 1  &  1  &  1  &  1 & $-1$   &  $-1$   &   $-i$  &   $i$    &  1   &   1  &  $1$  &  $1$      \\ \hline

$Z_3$   &   $\omega$     &    $\omega$  &   $\omega^2$   &   $\omega^2$   &   $\omega^2$ &   1  &  1   &  1  &   $\omega$  &   $\omega^2$ & $\omega^2$   &   $\omega$  &   $\omega^2$   &  $\omega$  \\ \hline

$U(1)_R$  &  2  &  2  &   2  &  2  &  2  & 2  &   1   &  1   &  1   & 1  &  1  &  1  &  1  & 1   \\

 \hline \hline

\end{tabular}
\caption{\label{tab:messenger} The driving fields, messenger fields and their transformation rules under the $S_4 \times Z_4\times Z_3$ and $U(1)_R$ symmetries.}
\end{table}

\subsection{The structure of the model}

The charged lepton sector is formulated at the renormalisable level with the introduction of three pairs of messengers $\Omega_i$ and $\Omega^{c}_i$ ($i=1,2,3$). Note that these messengers are chiral superfields with non-vanishing hypercharge $+2(-2)$ for $\Omega_i$ ($\Omega^{c}_i$). With the particles and their transformation properties listed in Tables \ref{tab:renormalisable} and \ref{tab:messenger}, we obtain the renormalisable superpotential for the charged leptons,
\begin{eqnarray}
\nonumber w_{\ell}&=&
z_1\left(L\Omega_3\right)_{\mathbf{1}}H_d
+z_2\left(\Omega^{c}_3\varphi_T\right)_{\mathbf{1}}\tau^c
+z_3\left(\left(\Omega^{c}_3\varphi_T\right)_{\mathbf{2}}\Omega_2\right)_{\mathbf{1}}
+z_4\left(\Omega^c_2\eta\right)_{\mathbf{1'}}\mu^c\\
&&+z_5\left(\left(\Omega^c_2\eta\right)_{\mathbf{2}}\Omega_1\right)_{\mathbf{1}}
\nonumber
+z_6\left(\Omega^c_1\eta\right)_{\mathbf{1}}e^c
+M_{\Omega_1}\left(\Omega_1\Omega^c_1\right)_{\mathbf{1}}
+z_7\Delta\left(\Omega_1\Omega^c_1\right)_{\mathbf{1}'}\\
&&+M_{\Omega_2}\left(\Omega_2\Omega^c_2\right)_{\mathbf{1}}
+z_8\Delta\left(\Omega_2\Omega^c_2\right)_{\mathbf{1}'}
+M_{\Omega_3}\left(\Omega_3\Omega^c_3\right)_{\mathbf{1}}\ ,
\end{eqnarray}
where general CP invariance again implies that all the order one coupling constants $z_i$ and the messenger masses $M_{\Omega_1}$, $M_{\Omega_2}$ and $M_{\Omega_3}$ are real. Furthermore, since the terms $\Delta\left(\Omega_1\Omega^c_1\right)_{\mathbf{1}'}$ and $\Delta\left(\Omega_2\Omega^c_2\right)_{\mathbf{1}'}$ lead to corrections to the $\Omega_1$ and $\Omega_2$  masses, respectively, and the mass scales of the messenger fields are much larger than the VEVs of the flavons, the contributions of these two operators can be safely neglected.

\begin{figure}[t!]
\begin{center}
\includegraphics[scale=.60]{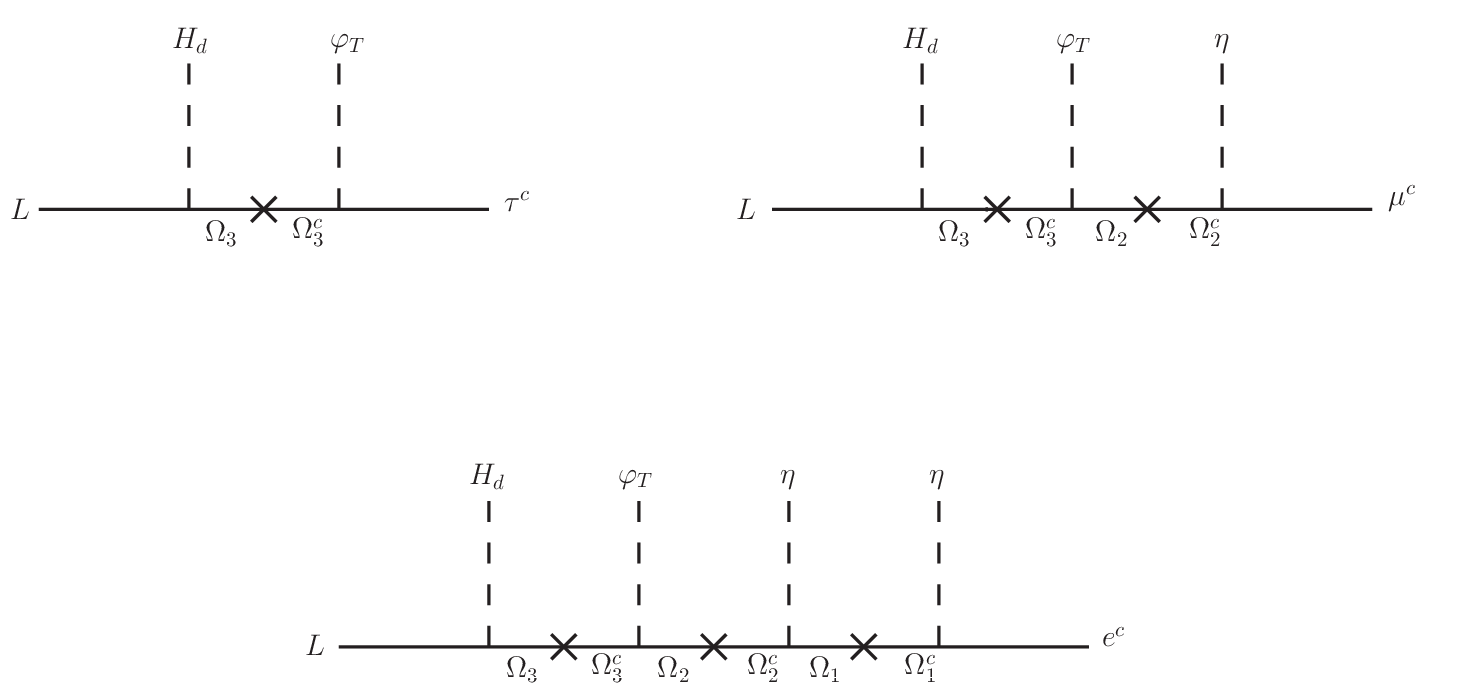}
\caption{\label{fig:charged_renor}The diagrams which generate the effective operators for the charged lepton masses, where crosses indicate the mass insertions for fermions.}
\end{center}
\end{figure}

Integrating out the messenger pairs $\Omega_i$ and $\Omega^c_i$ (the corresponding Feynman diagrams are shown in Fig. \ref{fig:charged_renor}), yields the following effective superpotential for the charged lepton masses:
\begin{eqnarray*}
w^{eff}_{\ell}&=&-\frac{z_1z_2}{M_{\Omega_3}}\;\left(L\varphi_T\right)_{\mathbf{1}}H_d\tau^c+\frac{z_1z_3z_4}{M_{\Omega_2}M_{\Omega_3}}\;\left(\left(L\varphi_T\right)_{\mathbf{2}}\eta\right)_{\mathbf{1}'}\mu^c\\
&&
-\frac{z_1z_3z_5z_6}{M_{\Omega_1}M_{\Omega_2}M_{\Omega_3}}\;\left(\left(L\varphi_T\right)_{\mathbf{2}}\left(\eta\eta\right)_{\mathbf{2}}\right)_{\mathbf{1}}H_de^c\ .
\end{eqnarray*}
Then, by applying the VEVs in Eq.~(\ref{eq:vacuum_charged_renor}), a diagonal charged lepton mass matrix is obtained with
\begin{eqnarray}
\label{eq:charged_mass_renor}
\hskip-0.07in m_{\tau}=-z_1z_2\frac{v_T}{M_{\Omega_3}}v_d ,\quad
m_{\mu}=z_1z_3z_4\frac{v_Tv_{\eta}}{M_{\Omega_2}M_{\Omega_3}}v_d,\quad
m_e=-z_1z_3z_5z_6\frac{v_Tv^2_{\eta}}{M_{\Omega_1}M_{\Omega_2}M_{\Omega_3}}v_d.
\end{eqnarray}
The mass hierarchies among the charged leptons are reproduced naturally without invoking another mechanisms. Here the VEVs of the flavons $\varphi_T$ and $\eta$ are responsible for the spontaneous breaking of both flavour and generalised CP symmetries. As the effective model in section~4, the $S_4$ flavour symmetry is broken completely and the remnant CP symmetry is $H^{\ell}_{CP}=\rho_{\mathbf{r}}(1)$ in the charged lepton sector.

Having completed the analysis of the charged lepton sector, we now turn to discuss the neutrino sector. We begin this by writing the renormalisable superpotential responsible for the light neutrino masses, which consists of the LO and relevant messenger terms:
\begin{equation}
w_{\nu}=w^{LO}_{\nu}+w^{\Sigma}_{\nu}\ ,
\end{equation}
where
\begin{eqnarray}
\nonumber
w^{LO}_{\nu}&=&y\left(LN^c\right)_{\mathbf{1}}H_u+y_1\left(\left(N^{c}N^c\right)_{\mathbf{3}'}\varphi_S\right)_{\mathbf{1}}+y_2\left(N^cN^c\right)_{\mathbf{1}}\xi+y_3\left(\left(N^cN^c\right)_{\mathbf{2}}\phi\right)_{\mathbf{1}}
\ ,\\
w^{\Sigma}_{\nu}&=&x_1\left(\left(N^c\Sigma\right)_{\mathbf{3}'}\varphi_S\right)_{\mathbf{1}}+x_2\left(\left(N^c\Sigma\right)_{\mathbf{2}}\phi\right)_{\mathbf{1}}+x_3\left(N^c\Sigma^c\right)_{\mathbf{1}'}\Delta+M_{\Sigma}\left(\Sigma\Sigma^c\right)_{\mathbf{1}}\ .~~
\end{eqnarray}
where the messenger field $\Sigma$ ($\Sigma^c$) is a chiral superfield carrying zero hypercharge, and all the parameters $x_i$ and $y_i$ are real due to generalised CP invariance. It is clear that the Dirac neutrino mass matrix takes the simple form
\begin{equation}
m_D=y v_u\left(\begin{array}{ccc}
1&0&0  \\
0&0&1 \\
0&1&0
\end{array}\right)\ .
\end{equation}
Further notice that $w^{LO}_{\nu}$ gives rise to the RH Majorana neutrino mass matrix $m^{LO}_M$, and after inserting the flavon VEVs of $\varphi_S$, $\phi$ and $\xi$ from Eq.~(\ref{eq:vacuum_neu_renor}), this is revealed to be
\begin{equation}
m^{LO}_M=y_1v_{s}\left(\begin{array}{ccc}
2& -1 & -1 \\
-1 &  2  &  -1  \\
-1 & -1  & 2
\end{array}\right)+y_2u\left(\begin{array}{ccc}
1 &0 &0 \\
0 &  0  & 1 \\
0 &  1 & 0
\end{array}\right)+y_3v_{\phi}\left(\begin{array}{ccc}
0  & 1  &  1  \\
1  &  1  &  0  \\
1  & 0  & 1
\end{array}\right).
\end{equation}
The resulting effective light neutrino mass matrix $m^{LO}_{\nu}=-m_D(m^{LO}_M)^{-1}m^{T}_D$ is exactly diagonalised by the tri-bimaximal mixing matrix $U_{TB}$,
\begin{equation}
U^{T}_{TB}m^{LO}_{\nu}U_{TB}=\text{diag}\left(m^{LO}_1,m^{LO}_2,m^{LO}_3\right)\ ,
\end{equation}
where the light neutrino masses $m^{LO}_{1,2,3}$ are
\begin{equation}
m^{LO}_1=-\frac{y^2v^2_u}{3y_1v_S+y_2u-y_3v_{\phi}},\;~
m^{LO}_2=-\frac{y^2v^2_u}{y_2u+2y_3v_{\phi}},\;~
m^{LO}_3=-\frac{y^2v^2_u}{3y_1v_S-y_2u+y_3v_{\phi}}\ .~~
\end{equation}
The reason why the tri-bimaximal mixing is produced is because the VEVs of $\varphi_S$, $\phi$ and $\xi$ preserve the Klein four subgroup generated by the tri-bimaximal $S$ and $U$ generators, as has been pointed out in section \ref{subsec:vacuum_renor}.
\begin{figure}[t!]
\begin{center}
\includegraphics[scale=.60]{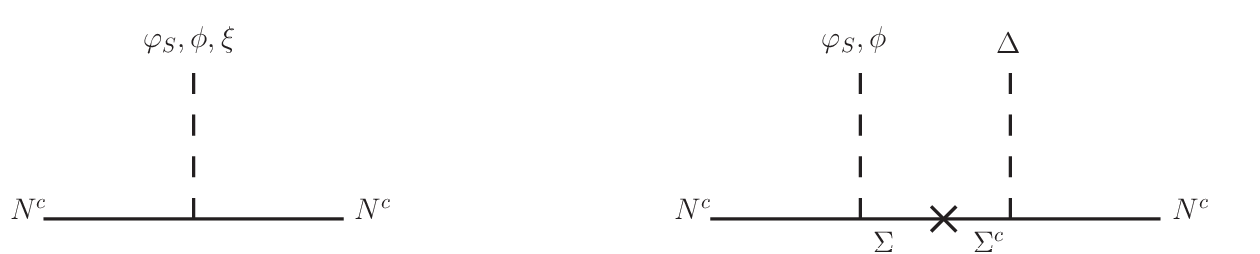}
\caption{\label{fig:neutrino_renor}The diagrams for the RH neutrino masses,
  where the cross indicates a fermionic mass insertion.}
\end{center}
\end{figure}
The LO and NLO contributions to the RH neutrino masses are shown diagrammatically in Fig.~\ref{fig:neutrino_renor}. Integrating out the messenger fields $\Sigma$ and $\Sigma^c$ yields the NLO effective operator
\begin{eqnarray}
\nonumber&&w^{NLO}_{\nu}=-\frac{x_2x_3}{M_{\Sigma}}\Delta\left(\left(N^cN^c\right)_{\mathbf{2}}\phi\right)_{\mathbf{1}'}\ .
\end{eqnarray}
The effective operator $\Delta\left(\left(N^cN^c\right)_{\mathbf{3}}\varphi_S\right)_{\mathbf{1}'}$ is not reproduced, since the contraction $N^cN^c$ vanishes from the antisymmetry of the associated $S_4$ Clebsch-Gordan coefficients, as is shown in Appendix \ref{app2}. Straightforwardly, we see that $w^{NLO}_{\nu}$ gives rise to a NLO contribution to $m_M$ of the form
\begin{eqnarray}
\nonumber&&m^{NLO}_M=x_2x_3\frac{v_{\Delta}v_{\phi}}{M_{\Sigma}}\left[\left(\begin{array}{ccc}
0  & 1  & -1 \\
1  & -1  &  0 \\
-1  & 0  & 1
\end{array}\right)
\right].
\end{eqnarray}
Note that the NLO contribution $m^{NLO}_M$ is induced by the flavon $\Delta$ which further breaks the remnant Klein four symmetry to  $Z^S_2$. Then, the RH neutrino mass matrix $m_M$ including the NLO contribution can be expressed as
\begin{eqnarray}
\nonumber m_M&\!=\!&m^{LO}_M+m^{NLO}_M\\
&\!=\!&\tilde{a}\begin{pmatrix}
  2&-1&-1\\-1&2&-1\\-1&-1&2\end{pmatrix}+\tilde{b} \begin{pmatrix}1&0&0\\0&0&1\\0&1&0 \end{pmatrix}+\tilde{c}\begin{pmatrix}0&1&1\\1&1&0\\1&0&1 \end{pmatrix}+\tilde{d}\begin{pmatrix}
    0&1&-1\\1&-1&0\\-1&0&1\end{pmatrix}\! ,~~~~~~
\end{eqnarray}
where the parameters $\tilde{a}$, $\tilde{b}$, $\tilde{c}$ and $\tilde{d}$ are defined as
\begin{equation}
\tilde{a}=y_1v_S,\quad \tilde{b}=y_2u,\quad \tilde{c}=y_3v_{\phi},\quad
\tilde{d}=x_2x_3\frac{v_{\Delta}v_{\phi}}{M_{\Sigma}}\ .
\end{equation}
The light neutrino mass matrix $m_{\nu}$ is given by the seesaw formula
\begin{eqnarray}
\nonumber m_{\nu}&\!=\!&-m_Dm^{-1}_Mm^{T}_D\\
\label{eq:neutrino_matrix_renor}&\!=\!&\alpha \begin{pmatrix} 2&-1&-1\\-1&2&-1\\-1&-1&2\end{pmatrix}
+\beta  \begin{pmatrix}1&0&0\\0&0&1\\0&1&0 \end{pmatrix}
+\gamma \begin{pmatrix}0&1&1\\1&1&0\\1&0&1 \end{pmatrix}
+\epsilon  \begin{pmatrix} 0&1&-1\\1&-1&0\\-1&0&1\end{pmatrix}\!.~~~~~~
\end{eqnarray}
It is the most general neutrino mass matrix consistent with the residual $Z^S_2$ flavour symmetry, as is shown in Eq.~\eqref{eq:S_neutrino}. The parameters $\alpha$, $\beta$, $\gamma$ and $\epsilon$ are given by,
\begin{eqnarray}
\nonumber&&\hspace{-10mm}\alpha=\frac{\tilde{a}}{-9 \tilde{a}^2+(\tilde{b}-\tilde{c})^2+3 \tilde{d}^2}\;,\quad \beta=-\frac{1}{3(\tilde{b}+2 \tilde{c})}+\frac{2 (\tilde{b}-\tilde{c})}{3\left[9\tilde{a}^2-(\tilde{b}-\tilde{c})^2-3\tilde{d}^2\right]}\;,\\
&&\hspace{-10mm}\gamma=-\frac{1}{3(\tilde{b}+2\tilde{c})}-\frac{\tilde{b}-\tilde{c}}{3\left[9\tilde{a}^2-(\tilde{b}-\tilde{c})^2-3\tilde{d}^2\right]}\;,\quad\epsilon=\frac{\tilde{d}}{-9 \tilde{a}^2+(\tilde{b}-\tilde{c})^2+3\tilde{d}^2}\;,
\end{eqnarray}
where the overall factor $y^2v^2_u$ has been omitted. We note that the first three terms in the light neutrino matrix of Eq.~\eqref{eq:neutrino_matrix_renor} preserve tri-bimaximal mixing while the last term, proportional to $\epsilon$, violates it. In the present model, the $\epsilon$ term is induced by the NLO contributions and hence suppressed by $v_{\Delta}/M_{\Sigma}$ with respect to $\alpha$, $\beta$ and  $\gamma$. This provides a natural explanation as to why the reactor angle as well as the deviations from maximal atmospheric mixing are relatively small. Although, their definite values cannot be predicted.

\begin{table}[t!]
\begin{center}
{\footnotesize\begin{tabular}{|c|c|c|c|c|c|}\hline\hline

  &   &    &     &     &   \\ [-0.15in]

&   &  $\alpha$ &   $\beta$   &   $\gamma$   &   $\epsilon$    \\ \hline

  &   &    &     &     &   \\ [-0.15in]

(i) & $\tilde{a}\in\mathbb{R},\,\tilde{b}\in\mathbb{R},\,\tilde{c}\in\mathbb{R},\,\tilde{d}\in\mathbb{R}$  &  real &  real &   real   &  real  \\ \hline

  &   &    &     &     &   \\ [-0.15in]

(ii) & $\tilde{a}\in\mathbb{C},\,\tilde{b}\in\mathbb{R},\,\tilde{c}\in\mathbb{R},\,\tilde{d}\in\mathbb{R}$ \text{with} $\tilde{a}=-\tilde{a}^{*}$  &  imaginary  &   real   &  real  &  real  \\ \hline

  &   &    &     &     &   \\ [-0.15in]

(iii) & $\tilde{a}\in\mathbb{R},\,\tilde{b}\in\mathbb{R},\,\tilde{c}\in\mathbb{R},\,\tilde{d}\in\mathbb{C}$ \text{with} $\tilde{d}=-\tilde{d}^{*}$  & real  &   real   &  real &  imaginary   \\  \hline

  &   &    &     &     &   \\ [-0.15in]

(iv) & $\tilde{a}\in\mathbb{C},\,\tilde{b}\in\mathbb{R},\,\tilde{c}\in\mathbb{R},\,\tilde{d}\in\mathbb{C}$ \text{with} $\tilde{a}=-\tilde{a}^{*}$ and $\tilde{d}=-\tilde{d}^{*}$  &  imaginary  &  real   &  real  & imaginary  \\ \hline \hline

\end{tabular}}
\caption{\label{tab:renormalisable_model}
The four allowed ranges of $\alpha$, $\beta$, $\gamma$ and $\epsilon$ as dictated by the domains of $\tilde{a}$, $\tilde{b}$, $\tilde{c}$ and $\tilde{d}$ in the renormalisable model.  }
\end{center}
\end{table}

From the vacuum alignment in section \ref{subsec:vacuum_renor}, we know that both $\tilde{b}$ and $\tilde{c}$ are real parameters, and $\tilde{a}$ and $\tilde{d}$ can be real or purely imaginary. Therefore, there are four possible cases allowed by the domains of the parameters in the renormalisable model, these are summarised in Table \ref{tab:renormalisable_model}. It turns out that each of these four cases will preserve different generalised CP transformations as well as lead to different phenomenological predictions. In what follows, we shall discuss the four cases in Table~\ref{tab:renormalisable_model}, one by one.

\begin{description}
\item[~~(i)]
In this case, both the triplet VEV $v_S $ and the singlet VEV $v_{\Delta}$ are real. The generalised CP symmetries for $g= {1}$ and $g=S$ are preserved in the neutrino sector, i.e. the residual CP symmetry in the neutrino sector is $H^{\nu}_{CP}=\left\{\rho_{\mathbf{r}}({1}),\rho_{\mathbf{r}}(S)\right\}$,
where $\mathbf{r}$ denotes the irreducible representations of $S_4$. For the
corresponding light neutrino mass matrix shown in Eq.~\eqref{eq:neutrino_matrix_renor}, the parameters $\alpha$, $\beta$, $\gamma$ and $\epsilon$ are all real, yielding a matrix that is precisely the same as the neutrino mass matrix of case (i) from the general analysis in section \ref{sec:general}. Therefore, the neutrino mixing matrix is of trimaximal form, and the predictions for lepton mixing angles and CP phases are given by Eqs.~(\ref{eq:i_angles-ph},\ref{eq:i_angles}). There is no CP violation, as the CP symmetry $X_{\mathbf{r}}=\rho_{\mathbf{r}}(1)$ is conserved in both the neutrino and charged lepton sectors. However this case
gives a viable description of the lepton mixing angles.

\item[~~(ii)]
In this case, $v_S$ is purely imaginary and $v_{\Delta}$ is real. The generalised CP symmetry is broken to $H^{\nu}_{CP}=\left\{\rho_{\mathbf{r}}(TST^2),\;\rho_{\mathbf{r}}(T^2ST)\right\}$ in the neutrino sector. The resulting parameter $\alpha$ is imaginary and $\beta$, $\gamma$ and $\epsilon$ are real. As a result, this is case (ii) discussed in section \ref{sec:general}, and the light neutrino masses are degenerate, i.e.  $|m_1|=|m_3|$.  Hence, this case is not viable.

\item[~~(iii)]
This case corresponds to the VEV $v_S$ being real and $v_{\Delta}$ being purely imaginary. Only two of the 24 generalised CP symmetries are preserved in the neutrino sector, i.e. $H^{\nu}_{CP}=\left\{\rho_{\mathbf{r}}(U),\;\rho_{\mathbf{r}}(SU)\right\}$.
Regarding the light neutrino mass matrix of Eq.~\eqref{eq:neutrino_matrix_renor}, the tri-bimaximal violating parameter $\epsilon$ is imaginary and $\alpha$, $\beta$ and $\gamma$ are real. Therefore, this case is identical to case (iii) of the general analysis inspired by symmetry arguments, and the predictions for its mixing angles and CP phases are as given in Eqs.~(\ref{eq:iii_anglesphases},\ref{eq:iii_angles}).  Notice that this case  produces maximal Dirac CP violation $|\delta_{CP}|=\pi/2$, which is the result of the mismatch between the remnant CP symmetries $H^{\ell}_{CP}=\left\{\rho_{\mathbf{r}}(1)\right\}$ and $H^{\nu}_{CP}=\left\{\rho_{\mathbf{r}}(U),\;\rho_{\mathbf{r}}(SU)\right\}$.
This case is therefore viable.

\item[~~(iv)]
Both the VEVs $v_S$ and $v_{\Delta}$ are purely imaginary in this case, the remnant CP symmetry in the neutrino sector turns out to be $H^{\nu}_{CP}=\left\{\rho_{\mathbf{r}}(TST^2U),\; \rho_{\mathbf{r}}(T^2STU)\right\}$. The parameters $\alpha$ and $\epsilon$ are imaginary while $\beta$ and $\gamma$ are real. Hence, the corresponding light neutrino mass matrix has the same form as that of case (iv) analysed in section \ref{sec:general}, and the resulting PMNS matrix is given in Eq.~\eqref{eq:iv_pmns} up to permutations of rows and columns. However, in the present model, the mass hierarchies among the three charged lepton are reproduced, as is shown in Eq.~\eqref{eq:charged_mass_renor}. Consequently, permuting the rows of the PMNS matrix is forbidden. Notice that even if we desired to exchange its columns, as the neutrino mass order is less constrained so far, we would always get $\sin^2\theta_{13}=1/3$, which is much larger than the experimental observations. Therefore, we conclude that this case is not viable for the renormalisable model.
\end{description}

In summary, in the renormalisable model there are only two viable cases, namely (i) with no CP violation, and (iii) with maximal CP violation. This is the same as in the effective model.

\section{\label{sec:concl}Conclusions}

The measurement of the reactor mixing angle, which is observed to be rather large, offers some encouragement that the measurement of leptonic CP violation, in particular the Dirac oscillation phase $\delta_{CP}$, may be possible in the not too distant future. This has led to renewed interest in theories that are able to predict the value of $\delta_{CP}$.

In this work, we have focused on a promising framework where (generalised) CP symmetry in the lepton sector is combined with a discrete family symmetry (and perhaps other symmetries). The CP and family symmetries are subsequently spontaneously broken, with different subgroups preserved in the neutrino and charged lepton sectors. The models presented in this paper provide a generalisation for the direct models reviewed in \cite{King:2013eh} to include CP symmetry.

Within this framework, we have addressed the question of spontaneous breaking of a generalised CP symmetry in particular for models based on an $S_4$ family symmetry. We have constructed two models of leptons based on $S_4$ family symmetry combined with a generalised CP symmetry $H_{CP}$, one at the effective level and another one at the renormalisable level. In both models, we have shown how the flavon potential can spontaneously break the symmetry $S_4 \rtimes H_{CP}$ down to $Z_2 \times H^{\nu}_{CP}$ in the neutrino sector. This symmetry breaking was simply assumed to happen in~\cite{Feruglio:2012cw} without any dynamical justification.

In our models, the choice of preserved CP symmetry $H^{\nu}_{CP}$ in the neutrino sector is controlled by free (real) parameters in the flavon potential, enabling us to dial the type of CP violation. Of the two realistic models of this kind that we have proposed, one at the effective level and another one at the renormalisable level, we find that both models predict trimaximal lepton mixing with CP being either fully preserved or maximally broken and the intermediate possibility forbidden by the structure of the models.
\vskip0.5cm

\textbf{Note Added:} After we submitted the present paper to the arXiv, a
related paper by the authors of~\cite{Feruglio:2012cw} appeared three days
later~\cite{Feruglio:2013hia}. Similar to our models, the proposed model of~\cite{Feruglio:2013hia} is formulated in a basis where the order three generator $T$ is diagonal, and the residual symmetry of the neutrino sector gives rise to trimaximal mixing. In contrast to our models which can accommodate any neutrino mass spectrum, the neutrino masses in the model of~\cite{Feruglio:2013hia} effectively depend on only two real parameters, thus it predicts a normal neutrino mass hierarchy as well as the absolute neutrino mass scale.

\section*{Acknowledgements}
The research was partially supported by  the National Natural Science Foundation of China under Grant Nos. 11275188 and 11179007, the EU ITN grants UNILHC 237920 and INVISIBLES 289442. SK and AJS acknowledge support from the STFC Consolidated  ST/J000396/1 grant. GJD and CL thank the School of Physics and Astronomy at the University of Southampton for hospitality.

\section*{Appendices}

\begin{appendix}

\section{\label{app1}Implications of a generalised CP symmetry}
\cleqn

It is well-known that at low energies there are three CP violating phases in the lepton sector: one of the Dirac type and two of the Majorana type. The strength of leptonic CP violation of the Dirac type, which can be observable through neutrino oscillations, is determined by the following CP-odd weak basis (WB) invariant \cite{Branco:1986gr}:
\begin{equation}
\label{eq:WB1}\mathcal{J}_1\equiv\mathrm{Tr}\left[h^{*}_{\nu},h_{l}\right]^3,\qquad \mathrm{with}\quad h_{\nu}=m_{\nu}m^{\dagger}_{\nu} \quad\mathrm{and}\quad h_{l}=m_{l}m^{\dagger}_l.
\end{equation}
It can be fully written in terms of physical observables as
\begin{eqnarray}
\mathcal{J}_1=-6\,i \,({m^2_{\mu}}-{m^2_e})\,({m^2_{\tau}}-{m^2_\mu})\,({m^2_{\tau}}-{m^2_e})\Delta m^2_{21}\,\Delta m^2_{31}\,\Delta m^2_{32}\,J_{CP},
\end{eqnarray}
where $\Delta m^2_{ij}=m^2_i-m^2_j$ are the light neutrino mass squared differences and the quantity $J_{CP}$ is the Jarlskog invariant \cite{Jarlskog:1985ht},
\begin{eqnarray}
\nonumber J_{CP}&=&Im\left[\left(U_{PMNS}\right)_{11}\left(U_{PMNS}\right)_{22}\left(U^{*}_{PMNS}\right)_{12}\left(U^{*}_{PMNS}\right)_{21}\right]\\
&=&\frac{1}{8}\sin(2\theta_{23})\sin(2\theta_{12})\sin(2\theta_{13})\cos\theta_{13}\sin\delta_{CP}.
\end{eqnarray}
where the PMNS matrix $U_{PMNS}$ has been parameterised as in Eq.~\eqref{eq:pmns_pdg}, $\theta_{ij}$ are the mixing angles chosen to lie in the first quadrant, and $\delta_{CP}$ is the Dirac CP-violating phase. It is obvious that $\mathcal{J}_1$ vanishes for $\delta_{CP}=0$, and vice versa the Dirac phase $\delta_{CP}$ would be zero if $\mathcal{J}_1$ vanishes. From Eq.~(\ref{eq:gc}), we see that the generalised CP invariance implies
\begin{equation}
h_{l}=X_Lh^{T}_lX^{\dagger}_L,\qquad h_{\nu}=X^{*}_Lh^{*}_{\nu}X^{T}_L .   \end{equation}
Therefore, we have
\begin{eqnarray}
\mathcal{J}_1=\mathrm{Tr}\left[h^{*}_{\nu},h_{l}\right]^3=\mathrm{Tr}\left(X_L\left[h_{\nu},h^{T}_{l}\right]^3X^{\dagger}_L\right)=\mathrm{Tr}\left[(h^{*}_{\nu})^{T},h^{T}_{l}\right]^3=-\mathrm{Tr}\left[h^{*}_{\nu},h_{l}\right]^3=0.
\end{eqnarray}
This indicates that there is no Dirac type CP violation if the generalised CP symmetry is preserved. In the case of Majorana neutrinos, there is also the possibility of the Majorana type CP violation. It has been established that the vanishing of the WB invariant $\mathcal{J}_1$ together with the following two WB invariants,
\begin{eqnarray}
\nonumber\mathcal{J}_2&=&\text{Im Tr}\left(h_lh^{*}_{\nu}m^{*}_{\nu}h^{*}_lm_{\nu}\right)\,, \\
\mathcal{J}_3&=&\text{Tr}\left[m_{\nu}h_lm^{*}_{\nu},h^{*}_l\right]^3 \,,
\end{eqnarray}
provides necessary and sufficient conditions for low energy CP invariance \cite{Branco:1998bw,Branco:2004hu,Dreiner:2007yz,Branco:2011zb}. Notice that $\mathcal{J}_2$ and $\mathcal{J}_3$ measure the Majorana type CP violation. Then, from the generalised CP invariance requirement of Eq.~(\ref{eq:gc}), we have
\begin{eqnarray}
&\mathcal{J}_2=\text{Im Tr}\left(X_Lh^{*}_lh_{\nu}m_{\nu}h_lm^{*}_{\nu}X^{\dagger}_L\right)=\text{Im Tr}\left(h_lh^{*}_{\nu}m^{*}_{\nu}h^{*}_lm_{\nu}\right)^{*}=0,\\\nonumber
&\mathcal{J}_3=\text{Tr}\left(X^{*}_L\left[m^{*}_{\nu}h^{*}_lm_{\nu},h_l\right]^3X^{T}_L\right)=\text{Tr}\left[(m_{\nu}h_lm^{*}_{\nu})^{T},(h^{*}_l)^{T}\right]^3=-\text{Tr}\left[m_{\nu}h_lm^{*}_{\nu},h^{*}_l\right]^3=0.
\end{eqnarray}
This means that CP will be conserved if the theory is invariant under a generalised CP transformation. As a result, the generalised CP symmetry must be spontaneously or explicitly broken in order to generate non-trivial CP phases.

\section{\label{app2}Group theory of $\bs{S_4}$}
\cleqn

$S_4$ is the permutation group of four distinct objects, and it is
isomorphic to the symmetry group of a regular octahedron. It has $4!=24$ elements and can be expressed in terms of two generators. However, in order to clearly see the connection to the groups $A_4$ and $S_3$ it is useful to express $S_4$ in terms of three generators, $S$, $T$ and $U$ \cite{Hagedorn:2010th}, which obey the multiplication rules \begin{eqnarray}\label{eq:presentation}
\begin{gathered}
 S^2 = T^3 = U^2=
 (S T)^3 = (S U)^2 = (T U)^2 =
  (S T U)^4 = 1 \ .
\end{gathered}
\end{eqnarray}
Notice that the generators $S$ and $T$ alone generate the group $A_4$, while the generators $T$ and $U$ alone generate the group $S_3$. Taking all possible combinations of $S$, $T$, and $U$ (subject to the rules of Eq.~(\ref{eq:presentation})), yields the 24 elements of $S_4$ which belong to 5 disjoint conjugacy classes. To emphasise the geometric aspect of $S_4$, we adopt Schoenflies notation, in which $k C_n$ designates a conjugacy class of $k$ elements that are all rotations by $\frac{2\pi}{n}$, to express the conjugacy classes of $S_4$ as
\begin{eqnarray}\nonumber
1 {C}_1&=& ~\{1\}\ ,  \\\nonumber
3 {C}_2&=&~\{S, TST^2, T^2ST\} \ , \\
6{C}_2^{\prime}&=& ~\{U, TU, SU, T^2U, STSU,ST^2SU \} \ , \\\nonumber
 8 {C}_3&=&~ \{T,ST,TS, STS, T^2, ST^2, T^2S, ST^2S\}   \ , \\\nonumber
 6C_4&=&~\{STU, TSU, T^2SU, ST^2U, TST^2U, T^2STU\}  \ .
\end{eqnarray}
These conjugacy classes can be used to deduce the various irreducible representations of $S_4$ because (by theorem) the number of the  irreducible representations must equal to the number of conjugacy classes. Thus, $S_4$ has five irreducible presentations. Additionally, the sum of the squares of the dimensions of the irreducible representations must equal the order of the group, \textit{i.e.} 24.  This implies the five irreducible representations of $S_4$ are two 1-dimensional ($\mathbf{1}$, $\mathbf{1}'$), one 2-dimensional ($\mathbf{2}$), and two 3-dimensional ($\mathbf{3}$ and $\mathbf{3}'$) irreducible representations. Our choice of the explicit basis for the representation matrices of $S$, $T$ and $U$ is listed in Table \ref{tab:representation} ~\cite{King:2011zj}.
\begin{table}[t!]
\begin{center}
\begin{tabular}{|c|c|c|c|}\hline\hline
 ~~  &  $S$  &   $T$    &  $U$  \\ \hline
~~~${\bf 1}$, ${\bf 1^\prime}$ ~~~ & 1   &  1  & $\pm1$  \\ \hline
   &   &    &    \\ [-0.16in]
${\bf 2}$ &  $\left( \begin{array}{cc}
    1&0 \\
    0&1
    \end{array} \right) $
    & $\left( \begin{array}{cc}
    \omega&0 \\
    0&\omega^2
    \end{array} \right) $
    & $\left( \begin{array}{cc}
    0&1 \\
    1&0
    \end{array} \right)$\\ [0.12in]\hline
   &   &    &    \\ [-0.16in]
${\bf 3}$, ${\bf 3^\prime}$ & $\frac{1}{3} \left(\begin{array}{ccc}
    -1& 2  & 2  \\
    2  & -1  & 2 \\
    2 & 2 & -1
    \end{array}\right)$
    & $\left( \begin{array}{ccc}
    1 & 0 & 0 \\
    0 & \omega^{2} & 0 \\
    0 & 0 & \omega
    \end{array}\right) $
    & $\mp\left( \begin{array}{ccc}
    1 & 0 & 0 \\
    0 & 0 & 1 \\
    0 & 1 & 0
    \end{array}\right)$
\\[0.22in] \hline\hline
\end{tabular}
\caption{\label{tab:representation}The $S_4$ representation matrices for the
  $S$, $T$ and $U$ elements in different irreducible representations, where $\omega=e^{2\pi i/3}$.}
\end{center}
\end{table}
Furthermore, the Kronecker products of these 5 different irreducible representations are
\begin{eqnarray}\nonumber
&&\bf 1 \otimes R = R\; , \;\;\; 1^\prime \otimes 1^\prime =1 \; , \;\;\;
  1^\prime \otimes 2 = 2 \;,\; \; \;  \bf 1^\prime \otimes 3 = 3^\prime \; ,
  \;\;\; 1^\prime \otimes 3^\prime = 3  \;, \\[0mm]
\nonumber
&&\bf 2 \otimes 2 = (1\oplus  2)_s \oplus 1^\prime_a\; , \;\;\;
2 \otimes 3 = 2 \otimes 3^\prime = 3 \oplus 3^\prime\; ,\\[0mm]
&&\bf 3 \otimes 3 = 3^\prime \otimes 3^\prime = (1\oplus 2 \oplus 3^\prime)_s
\oplus 3_a \; , \;\;\;
3 \otimes 3^\prime = 1^\prime \oplus 2 \oplus 3 \oplus 3^\prime \; ,
\end{eqnarray}
where ${\bf{R}}$ stands for any $S_4$ representation, and the index $s$ ($a$) denotes symmetric (antisymmetric) combinations. These Kronecker products, along with the explicit forms of the generators in Table \ref{tab:representation} can be used to calculate the corresponding Clebsch-Gordan (CG) coefficients. These CG coefficients can be also be found
in other works, e.g. Ref.~\cite{King:2011zj}, but we list them here for completeness. In the following reporting of the CG coefficients of $S_4$, we
use $\alpha_i$ to denote the elements of the first representation and $\beta_j$ to indicate those of the second representation of the product.
Furthermore, ``$n$'' counts the number of ``primes'' in the Kronecker product (e.g. in $\bf{1}\otimes \bf{1'}=\bf{1'}$, $n=2$).
$$
\begin{array}{ccc}
{\bf 1}^{(\prime)} \otimes {\bf 1}^{(\prime)} ~\rightarrow ~{\bf
  1}^{(\prime)} ~~
\left\{ \begin{array}{c}
~\\n=\mathrm{even}\\~
\end{array}\right.
&
\left.
\begin{array}{c}
{\bf 1}^{\phantom{\prime}} \otimes {\bf 1}^{\phantom{\prime}} ~\rightarrow ~{\bf 1}^{\phantom{\prime}}\\
{\bf 1}^{{\prime}} \otimes {\bf 1}^{{\prime}} ~\rightarrow ~{\bf 1}^{\phantom{\prime}}\\
{\bf 1}^{\phantom{\prime}} \otimes {\bf 1}^{{\prime}} ~\rightarrow ~{\bf 1}^{{\prime}}
\end{array}
\right\}
&
\alpha \beta \ ,
\\[10mm]
{\bf 1}^{(\prime)} \otimes \;{\bf 2} \;~\rightarrow \;~{\bf 2}^{\phantom{(\prime)}}~~ \left\{
\begin{array}{c}
n=\mathrm{even} \\
n=\mathrm{odd}
\end{array} \right.
&
\left.
\begin{array}{c}
{\bf 1}^{\phantom{\prime}} \otimes {\bf 2} ~\rightarrow ~{\bf 2} \\
{\bf 1}^{\prime} \otimes {\bf 2} ~\rightarrow ~{\bf 2}\\
\end{array}\;~
\right\}
&
 \alpha \begin{pmatrix} \beta_1 \\ (-1)^n \beta_2\end{pmatrix}  ,
\\[7mm]
{\bf 1}^{(\prime)} \otimes {\bf 3}^{(\prime)} ~\rightarrow ~{\bf 3}^{(\prime)}
~~ \left\{ \begin{array}{c}
~\\[3mm]n=\mathrm{even} \\[3mm]~
\end{array}\right.
&
\left.
\begin{array}{c}
{\bf 1}^{\phantom{\prime}} \otimes {\bf 3}^{\phantom{\prime}} ~\rightarrow ~{\bf 3}^{\phantom{\prime}}
\\
{\bf 1}^{{\prime}} \otimes {\bf 3}^{{\prime}} ~\rightarrow ~{\bf 3}^{\phantom{\prime}}
\\
{\bf 1}^{\phantom{\prime}} \otimes {\bf 3}^{{\prime}} ~\rightarrow ~{\bf 3}^{{\prime}}
\\
{\bf 1}^{{\prime}} \otimes {\bf 3}^{\phantom{\prime}} ~\rightarrow ~{\bf 3}^{{\prime}}
\end{array}
\right\}
&
 \alpha   \begin{pmatrix} \beta_1 \\  \beta_2\\\beta_3 \end{pmatrix}  ,
\\[12.2mm]
{\bf 2} \;\; \otimes \;\;{\bf 2} \;~\rightarrow \;~{\bf 1}^{(\prime)} ~~ \left\{\begin{array}{c}
n=\mathrm{even}\\
n=\mathrm{odd}
\end{array}\right.
&
\left.
\begin{array}{c}
{\bf 2} \otimes {\bf 2} ~\rightarrow ~{\bf 1}^{\phantom{\prime}} \\
{\bf 2} \otimes {\bf 2} ~\rightarrow ~{\bf 1}^{{\prime}}
\end{array}~\;
\right\}
&
 \alpha_1 \beta_2 + (-1)^n \alpha_2 \beta_1 \ ,\\\\
{\bf 2} \;\;\otimes \;\; {\bf 2} ~\;\rightarrow \;~{\bf 2}^{\phantom{(\prime)}} ~~ \left\{ \begin{array}{c}
~\\[-3mm] n=\mathrm{even}\\[-3mm]~
\end{array}\right.
&
\left.
\begin{array}{c}
~\\[-3mm]
{\bf 2} \otimes {\bf 2} ~\rightarrow ~{\bf 2} \\[-3mm]~
\end{array}~~\,
\right\}
&
\begin{pmatrix} \alpha_2 \beta_2 \\  \alpha_1\beta_1 \end{pmatrix} ,
\\[6mm]
{\bf 2}\;\; \otimes \; {\bf 3}^{{(\prime)}} ~\rightarrow ~{\bf 3}^{{(\prime)}} ~~ \left\{\begin{array}{c}
~\\[-2mm] n=\mathrm{even}\\ \\[2mm]
n=\mathrm{odd}\\[-2mm]~
\end{array}\right.
&
\left.
\begin{array}{c}
{\bf 2} \otimes {\bf 3}^{\phantom{\prime}} ~\rightarrow ~{\bf 3}^{\phantom{\prime}} \\
{\bf 2} \otimes {\bf 3}^{{\prime}} ~\rightarrow ~{\bf 3}^{{\prime}} \\[3mm]
{\bf 2} \otimes {\bf 3}^{\phantom{\prime}} ~\rightarrow ~{\bf 3}^{{\prime}} \\
{\bf 2} \otimes {\bf 3}^{{\prime}} ~\rightarrow ~{\bf 3}^{\phantom{\prime}}
\end{array}\;
\right\}
&
 \alpha_1 \begin{pmatrix} \beta_2 \\  \beta_3\\\beta_1 \end{pmatrix} + (-1)^n
\alpha_2 \begin{pmatrix} \beta_3 \\  \beta_1\\\beta_2 \end{pmatrix}  ,
\\[13.5mm]
{\bf 3}^{(\prime)} \otimes {\bf 3}^{(\prime)} ~\rightarrow ~{\bf 1}^{(\prime)}
~~ \left\{ \begin{array}{c}
~\\n=\mathrm{even}\\~
\end{array}\right.
&
\left.\begin{array}{c}
{\bf 3}^{\phantom{\prime}} \otimes {\bf 3}^{\phantom{\prime}} ~\rightarrow ~{\bf 1}^{\phantom{\prime}}
\\
{\bf 3}^{{\prime}} \otimes {\bf 3}^{{\prime}} ~\rightarrow ~{\bf 1}^{\phantom{\prime}}
\\
{\bf 3}^{\phantom{\prime}} \otimes {\bf 3}^{{\prime}} ~\rightarrow ~{\bf 1}^{{\prime}}
\end{array}\right\}
&
 \alpha_1 \beta_1 +\alpha_2\beta_3+\alpha_3\beta_2 \ ,
\\[9mm]
{\bf 3}^{(\prime)} \otimes {\bf 3}^{(\prime)} ~\rightarrow ~{\bf 2}^{\phantom{(\prime)}} ~~
\left\{ \begin{array}{c}
~\\[-3mm]
n=\mathrm{even}\\ \\[1mm]
n=\mathrm{odd}\\[-4.5mm]~
\end{array}\right.
&
\left.\begin{array}{c}
{\bf 3}^{\phantom{\prime}} \otimes {\bf 3}^{\phantom{\prime}} ~\rightarrow ~{\bf 2} \\
{\bf 3}^{{\prime}} \otimes {\bf 3}^{{\prime}} ~\rightarrow ~{\bf 2} \\[3mm]
{\bf 3}^{\phantom{\prime}} \otimes {\bf 3}^{{\prime}} ~\rightarrow ~{\bf 2} \\
\end{array}\;
\right\}
&
\begin{pmatrix} \alpha_2 \beta_2 +\alpha_3 \beta_1+\alpha_1\beta_3\\
(-1)^n(\alpha_3 \beta_3+\alpha_1\beta_2+\alpha_2\beta_1) \end{pmatrix} ,
\\
\end{array}
$$
$$
\begin{array}{ccc}
\hskip-0.35in{\bf 3}^{(\prime)} \otimes {\bf 3}^{(\prime)} ~\rightarrow ~{\bf 3}^{(\prime)}
~~ \left\{\begin{array}{c}
~\\n=\mathrm{odd}\\~
\end{array}\right.
&
\left.\begin{array}{c}
{\bf 3}^{\phantom{\prime}} \otimes {\bf 3}^{\phantom{\prime}} ~\rightarrow ~{\bf 3}^{{\prime}}
\\
{\bf 3}^{\phantom{\prime}} \otimes {\bf 3}^{{\prime}} ~\rightarrow ~{\bf 3}^{\phantom{\prime}}
\\
{\bf 3}^{{\prime}} \otimes {\bf 3}^{{\prime}} ~\rightarrow ~{\bf 3}^{{\prime}}
\end{array}\right\}
&
\begin{pmatrix}
2 \alpha_1 \beta_1-\alpha_2\beta_3-\alpha_3\beta_2 \\
2 \alpha_3 \beta_3-\alpha_1\beta_2-\alpha_2\beta_1 \\
2 \alpha_2 \beta_2-\alpha_3\beta_1-\alpha_1\beta_3
 \end{pmatrix} ,
\\[9mm]
\hskip-0.35in {\bf 3}^{(\prime)} \otimes {\bf 3}^{(\prime)} ~\rightarrow ~{\bf 3}^{(\prime)}~~\left\{\begin{array}{c}
~\\n=\mathrm{even}\\~
\end{array}\right.
&
\left.\begin{array}{c}
{\bf 3}^{\phantom{\prime}} \otimes {\bf 3}^{\phantom{\prime}} ~\rightarrow ~{\bf 3}^{\phantom{\prime}}
\\
{\bf 3}^{{\prime}} \otimes {\bf 3}^{{\prime}} ~\rightarrow ~{\bf 3}^{\phantom{\prime}}
\\
{\bf 3}^{\phantom{\prime}} \otimes {\bf 3}^{{\prime}} ~\rightarrow ~{\bf 3}^{{\prime}}
\end{array}\right\}
&
\begin{pmatrix}
\alpha_2\beta_3-\alpha_3\beta_2 \\
\alpha_1\beta_2-\alpha_2\beta_1 \\
\alpha_3\beta_1-\alpha_1\beta_3
 \end{pmatrix}  .
\end{array}\\[3mm]
$$
Now that we have discussed the simpler aspects of the group theory of $S_4$, we turn to the more complex topic of the automorphisms of the group $S_4$, i.e. the generalised CP transformations $X_{\bf r}$.

Although the most general solution for the $X_{\bf r}$ is given by the representation $\rho_{\bf r}(h)$ of the $S_4$ group elements $h\in S_4$, as is claimed in Ref.~\cite{Holthausen:2012dk}, it is still instructive to see how this explicitly occurs. Let us first consider the consistency equation Eq.~(\ref{eq:consistency}) for the generators $S$, $T$ and $U$ in the faithful representations $\mathbf{3}$ and $\mathbf{3}'$:
\begin{eqnarray}
\nonumber&&X_{\mathbf{3}(\mathbf{3}^{\prime})}\rho^{*}_{\mathbf{3}(\mathbf{3}^{\prime})}(S)X^{-1}_{\mathbf{3}(\mathbf{3}^{\prime})}=\rho_{\mathbf{3}(\mathbf{3}^{\prime})}(S^{\prime}), \\\label{eq:appconsist}
&&X_{\mathbf{3}(\mathbf{3}^{\prime})}\rho^{*}_{\mathbf{3}(\mathbf{3}^{\prime})}(T)X^{-1}_{\mathbf{3}(\mathbf{3}^{\prime})}=\rho_{\mathbf{3}(\mathbf{3}^{\prime})}(T^{\prime}),\\\nonumber
&&X_{\mathbf{3}(\mathbf{3}^{\prime})}\rho^{*}_{\mathbf{3}(\mathbf{3}^{\prime})}(U)X^{-1}_{\mathbf{3}(\mathbf{3}^{\prime})}=\rho_{\mathbf{3}(\mathbf{3}^{\prime})}(U^{\prime}).
\end{eqnarray}
It is clear from the above equations that the orders of primed and unprimed generators  must be identical, i.e. $S^{\prime}$ and $U^{\prime}$ must be order 2 elements and $T'$ an order 3 element. As a result, the consistency equations constrain the mappings (automorphisms) such that the generators $S,T,U$ can only be mapped into specific (unions of) conjugacy classes. Namely,
\be
S^{\prime} \in 3C_2 \cup 6 C_2^{\prime} \ , \qquad
T^{\prime}\in 8 C_3 \ , \qquad
U^{\prime}\in 3C_2 \cup 6 C_2^{\prime}  \ .
\ee
Additional constraints can be derived by considering the singlet representation ${\bf 1^{\prime}}$. Since $S=T=1$ and $U=-1$, $S^{\prime}$ cannot be in the conjugacy class $6C_2^{\prime}$. Likewise the element $U'$ cannot be in the conjugacy class $3C_2$. This leaves us the automorphisms

\begin{equation}
S^{\prime}\in 3C_2 \ , \qquad
T^{\prime}\in  8 C_3 \ , \qquad
U^{\prime}\in  6 C_2^{\prime} \ .
\end{equation}
By investigating all  $3\times 8\times 6=144$ possible values of $S^{\prime}$, $T^{\prime}$ and $U^{\prime}$, we find only 24 solutions that satisfy the corresponding consistency equations Eq.~(\ref{eq:appconsist}), which can be compactly written as
\begin{equation}
\label{eq:CPtrans_3}X_{\mathbf{3}(\mathbf{3}')}=\rho_{\mathbf{3}(\mathbf{3}')}(h),\qquad
h\in S_4 \ ,
\end{equation}
with
\begin{eqnarray}
\nonumber&& X_{\mathbf{3}(\mathbf{3}')}\rho^{*}_{\mathbf{3}(\mathbf{3}')}(S)X^{-1}_{\mathbf{3}(\mathbf{3}')}=\rho_{\mathbf{3}(\mathbf{3}')}(hSh^{-1}),\\ \nonumber&&X_{\mathbf{3}(\mathbf{3}')}\rho^{*}_{\mathbf{3}(\mathbf{3}')}(T)X^{-1}_{\mathbf{3}(\mathbf{3}')}=\rho_{\mathbf{3}(\mathbf{3}')}(hT^2h^{-1}), \\ &&X_{\mathbf{3}(\mathbf{3}')}\rho^{*}_{\mathbf{3}(\mathbf{3}')}(U)X^{-1}_{\mathbf{3}(\mathbf{3}')}=\rho_{\mathbf{3}(\mathbf{3}')}(hUh^{-1}).
\end{eqnarray}
Continuing to the $\mathbf{2}$-dimensional irreducible representation, we have
\begin{equation}
\rho^{*}_{\mathbf{2}}(S)=\rho_{\mathbf{2}}(S),\qquad \rho^{*}_{\mathbf{2}}(T)=\rho_{\mathbf{2}}(T^2),\qquad \rho^{*}_{\mathbf{2}}(U)=\rho_{\mathbf{2}}(U).
\end{equation}
Thus, the generalised CP transformations consistent with the faithful three dimensional irreducible representations are
\begin{equation}
X_{\mathbf{2}}=\rho_{2}(h),\qquad h\in S_4\ .
\end{equation}
Finally, for the singlet representations $\mathbf{1}$ and $\mathbf{1}'$, we take
\begin{equation}
X_{\mathbf{1}(\mathbf{1}')}=\rho_{\mathbf{1}(\mathbf{1}')}(h),\qquad h\in
S_4\ .
\end{equation}
Therefore the generalised CP transformation consistent with an $S_4$ family symmetry is of the same form as the flavour group transformation, i.e. \begin{equation}
X_{\bf r} ~=~\rho_{\bf r} (h) \ , \qquad h\in S_4 \ .
\end{equation}
This provides a proof to the statement in Ref.\cite{Holthausen:2012dk} that the generalised CP transformation group consistent with a $S_4$ flavour group is $S_4$.\footnote{Said again, the generalised CP transformation group is the identity up to inner automorphisms.}

We end this appendix by noting that our basis for $S$, $T$ and $U$ is related to the basis choice in Ref.~\cite{Feruglio:2012cw} (denoted here as  $\widetilde{S}$, $\widetilde{T}$ and $\widetilde{U}$)  by the unitary transformation
\begin{equation}
V=\left(\begin{array}{ccc}
\sqrt{\frac{2}{3}}  &   \frac{1}{\sqrt{3}}  &  0  \\
-\frac{1}{\sqrt{6}}  &  \frac{1}{\sqrt{3}}  & -\frac{i}{\sqrt{2}} \\
-\frac{1}{\sqrt{6}}  &  \frac{1}{\sqrt{3}}  & \frac{i}{\sqrt{2}}
\end{array}\right),
\end{equation}
such that (for the $\mathbf{3}'$ representation)
\be
\nonumber
\widetilde{S}=V^{\dagger}SV=\left(
\begin{array}{ccc}
 -1 & 0 & 0 \\
 0 & 1 & 0 \\
 0 & 0 & -1
\end{array}
\right),\qquad
\widetilde{T}=V^{\dagger}TV=\frac{1}{2}\left(
\begin{array}{ccc}
 1 & \sqrt{2} & 1 \\
 \sqrt{2} & 0 & -\sqrt{2} \\
 -1 & \sqrt{2} & -1
\end{array}
\right),
\ee
\be
\widetilde{U}=V^{\dagger}UV=\left(
\begin{array}{ccc}
 1 & 0 & 0 \\
 0 & 1 & 0 \\
 0 & 0 & -1
\end{array}
\right),
\ee
and the generalised CP transformations in the two bases are related by \begin{equation}
\widetilde{X}_{\mathbf{3}(\mathbf{3}')}=V^{\dagger}X_{\mathbf{3}(\mathbf{3}')}V^{*}.
\end{equation}
It is interesting to note that the matrix $V$ satisfies
\begin{equation}
V^{\dagger}V^{*}=\left(
\begin{array}{ccc}
 1 & 0 & 0 \\
 0 & 1 & 0 \\
 0 & 0 & -1
\end{array}
\right) =\widetilde{U} ,
\end{equation}
so that $\wt X_{\bf 3(3')} = V^\dagger X_{\bf 3(3')} V \wt U$  is explicitly an element of $S_4$ in the basis with the tilde, given that $X_{\bf 3(3')}$ is an element of $S_4$ in our basis.

\end{appendix}

\end{document}